\documentclass[reprint,amsmath,amssymb,aps]{revtex4-1}
\usepackage{epsfig}
\usepackage{graphicx}
\usepackage{dcolumn}
\usepackage{comment}
\usepackage{color}
\usepackage{ulem}
\usepackage{longtable}
\usepackage{amsthm,amsmath}

\begin{document}

\title{Two-dipole and three-dipole dispersion coefficients for interaction of alkaline-earth atoms with alkaline-earth atoms and alkaline-earth ions}

\author{Neelam Shukla$^1$, Bindiya Arora$^2$\footnote{bindiya.phy@gndu.ac.in},  Lalita Sharma$^1$, and Rajesh Srivastava$^1$\footnote{rajesh.srivastava@ph.iitr.ac.in}}
\affiliation{$^1$Department of Physics, Indian Institute of Technology Roorkee, 247667, Roorkee, India}
\affiliation{$^2$Department of Physics, Guru Nanak Dev University, Amritsar, Punjab-143005, India}

\begin{abstract}
Apropos to the growing interest in the study of long-range interactions which for their applications in cold atom physics, we have performed theoretical calculation for the two-dipole $C_6$ and three-dipole $C_9$ dispersion coefficients involving alkaline-earth atoms with alkaline-earth atoms and alkaline-earth ions. The $C_6$ and $C_9$ coefficients are expressed in terms of the dynamic dipole polarizabilities, which are calculated  using relativistic methods. Thereafter, the calculated $C_6$ coefficients for the considered alkaline-earth atoms among themselves are compared with the previously reported values. Due to unavailability of any other earlier theoretical or experimental results, for the $C_6$ coefficients for  alkaline-earth atoms with alkaline-earth ions and the  $C_9$ coefficients, we have performed separate fitting calculations and compared. Our calculations match in an excellent manner with the fitting calculations. We have also reported the oscillator strengths for the leading transitions and static dipole polarizabilities for the ground states of the alkaline-earth ions, i.e., Mg$^+$, Ca$^+$, Sr$^+$, and Ba$^+$  as well as  the alkaline-earth atoms, i.e., Mg, Ca, Sr, and Ba. These, when compared with the available experimental results, show good agreement.
\end{abstract}

\maketitle
\section{Introduction}
Laser cooled atoms and ions have been the subject of several recent investigations covering many aspects of ultra-cold atom physics. Accurate knowledge of long-range part of the interatomic interaction between cooled atoms and ions can be viewed as the cornerstone of research on the atom-ion hybrid traps \cite{smith2005,zipkes2010}, experiments on photoassociation \cite{aymar2011,gacesa2016}, determination of scattering lengths, fluorescence spectroscopy, analysis of Feshbach resonances, probing extra dimensions to accommodate Newtonian gravity in quantum mechanics, determination of stability of Bose-Einstein condensates, and many more \cite{roberts1998,amiot2000,leo2000,leanhardt2003,yuju2004}. The study of long-range interaction is of special interest for low-temperature collisions \cite{zipkes2010,schmid2010,bodo2008,zhang2009,zhang20091,idziaszek2009,gao2010,cote2000,cote20001,makarov2003} and is quite important in the determination of collisional frequency shift. In various reports, it has been proposed that controlled ion-atom cold collisions can be used for future quantum information processing \cite{doerk2010,harter2014}. Further, the investigation of the long-range interaction yields vital information which helps to understand the different aspects of ion-atom bound state \cite{cote2002}, charge transfer processes \cite{cote2000,cote20001,zhang2009,zhang20091,sayfutyarova2013}, spin exchange reactions 
and formation of cold molecular ions \cite{schneider2010,staanum2010}.

The long-range interaction potential is mostly expressed in power series of the inverse of the interatomic separation $R$. The leading $R^{-6}$ term in this series representation is called the dispersion term, with coefficient $C_6$ which is of particular interest when the two atomic species are considered. On the other hand, the term with coefficient $C_9$ is regarded as the main contributor to the non-additive part of the interaction energy among three atomic species. The first theoretical study on ion-atom interactions goes back to the approach given by Langevin~\cite{langevin} and Margenau~\cite{marg}. Since then, a number of theoretical approaches have been reported. In a recent review, Koutselos and Mason \cite{koutselos1986} summarized data on ion-atom dispersion $C_6$ coefficients using Slater-Kirkwood formula. Ahlrichs \textit{et al.} \cite{ahlrichs1988} also used the same method to calculate dispersion coefficients for Li$^+$, Na$^+$, K$^+$, P$^-$, and Cl$^-$  ions interacting with He, Ne, and  Ar atoms and found good agreement between the experimental as well as theoretical results. Koutselos \textit{et al.} \cite{koutselos1990} determined interaction potentials from the universal interaction curves between the noble gas--noble gas, alkali ion--noble gas, and halogen ion--noble gas interactions which agree well with the experimentally determined potentials. They also successfully reproduced the measured diffusion coefficients and ion mobilities. Mitroy and Zhang \cite{mitroy2008} calculated long-range dispersion coefficients for Mg$^+$ and Ca$^+$  interacting with a number of atoms by using sum rule. Tang \textit{et al.} \cite{tang2010} evaluated $C_6$ coefficients for the ground and excited states of Li, Li$^+$, and Be$^+$ interacting with the He, Ne, Ar, Kr, and Xe  atoms in their ground states. They used variational Hylleraas method to determine the necessary list of multipole matrix elements. Sukhjit \textit{et al.} \cite{singh2016} reported the long-range dispersion coefficients for the interaction of inert gas atoms with the alkali atoms as well as alkaline-earth and alkali ions. They employed relativistic coupled-cluster method to estimate the dynamic dipole and quadrupole polarizabilities of the alkali atoms and singly ionized alkaline-earth atoms, whereas a relativistic random phase approximation approach was taken to determine these quantities for the closed-shell configured inert gas atom and singly-ionized alkali and doubly ionized alkaline-earth ions. The accuracy of these results was checked by comparing their static polarizability values with the experimental results. Tim and Toma \cite{gould2016c} obtained dispersion coefficient between rare gas atom and ions by implementing time-dependent density functional theory (TDDFT) with exchange kernels. They first calculated frequency-dependent dipole polarizabilities for atoms and ions and then integrated them over frequency to produce $C_6$ coefficients. 

In addition to above mentioned theoretical studies for long-range ion-atom interaction, there are several experimental studies which have been performed on ion-atom interactions to understand the nature of force between ultracold atoms and ions in hybrid ion-atom traps and some of these are described here.  Smith \textit{et al.} \cite{smith2005} successfully constructed a hybrid ion-atom trap which was designed to co-trap laser-cooled Ca$^+$ ions along with cold Na atoms. The first experiment on ion-atom cold collisions was reported by Grier \textit{et al.} \cite{grier2009} for an alkaline-earth like system viz. Yb +Yb$^+$ at energies ranging from 35 mK to 45 K. Next, heteronuclear Yb$^+$-Ca collisions were investigated by Zipkes \textit{et al.} \cite{zipkes2010cold, zipkes2010trapped} in the range of 0.2--5 K.  Hall \textit{et al.} \cite{hall2011light} and Hall \textit{et al.} \cite{hall2012millikelvin} performed an experiment on cold reactive collisions among laser-cooled  ions and atoms. 
Sullivan \textit{et al.} \cite{sullivan2012} studied the collision of  Ba$^+$ ions with Ca atoms, in which the charge exchange process is energetically prohibited unless Ca is electronically excited by the cooling laser. Ravi \textit{et al.} \cite{ravi2012} investigated the cooling of the Rb$^+$ ion by Rb atoms, where they experimentally demonstrated that rubidium ions cool in contact with magneto-optically trapped rubidium atoms, unlike the general expectation of ion heating. The cooling process is described theoretically and justified with numerical simulations, which involves resonant charge exchange collisions. Lee, Ravi and Rangwala \cite{lee2013} also investigated the same system, i.e., Rb$^+$ + Rb, where the ions were produced directly from the atomic cloud by two photon ionization. The use of such an alkaline-earth molecular, ionic system helps to examine the distinct processes that are essential for quantum information storage \cite{tacconi2011,mclaughlin2014}.

 Besides the two-body interaction, the three-body interaction has also been of immense interest for the past few decades. The first few studies on three-dipole interaction were by Axilrod \& Teller \cite{axilrod1943} and Muto \cite{muto1943}. Axilrod \& Teller \cite{axilrod1943} predicted the order of magnitude of the  $C_9$ coefficient and Muto \cite{muto1943} evaluated its value by using a simple atomic model. Later, Axilrod \cite{axilrod1951} also used a simplified atomic model to compute  $C_9$ coefficients, which were in agreement with those of Muto \cite{muto1943}. Further, Marinescu and Starace \cite{marinescu1997} investigated the non-additive part of the long-range interaction by implementing non-degenerate perturbation theory up to the third order and calculated the dispersion coefficients $C_9$ for three alkali-metal atoms interacting through their electric dipole moments. Both the homonuclear and heteronuclear cases were studied in this work.  Patil \& Tang \cite{patil1997} studied two- and three-body dispersion coefficients for alkali isoelectronic sequence. They calculated multipolar matrix elements by using simple wave-functions which were based on an asymptotic behaviour and the binding energies of the valence electron. These matrix elements allowed them to evaluate polarizabilities and dispersion coefficients of heteronuclear and homonuclear interactions from $C_6$ to C$_{24}$. Lilienfeld \& Tkatchenko \cite{anatole2010} presented a numerical estimation of the leading two- and three-body dispersion energy terms in van der Waals interactions for a broad range of molecules and solids. These calculations were based on Axilrod-Teller-Muto and London expressions where the required dispersion coefficients $C_6$ and $C_9$, are evaluated from the electron density. These coefficients were investigated by Huanga \& Sun \cite{huang2011} using a variational stable method of Gao \& Starace \cite{gao1989}, as well as the simple ground state wave-function of the valence electron previously suggested by Patil \& Tang \cite{patil1997}.

Ultracold alkaline-earth atoms are widely used in the precision measurements and quantum simulation studies. Due to their unique atomic structure, they can be used to investigate the quantum many-body system problems, such as Kondo and SU(N) physics, the simulation of synthetic gauge fields, etc. However, to thoroughly explore the potential of ultracold alkaline-earth atoms, these systems need to be studied in detail \cite{zhang2020,zhang2019}. Interestingly, recent developments in the experimental methods have opened the way for combining ultracold trapped ions and atoms in a single experimental setup \cite{wineland2013,haroche2013}. Most of the ion-atom experiments use alkaline-earth ions trapped and laser-cooled in a Paul trap immersed into ultracold neutral alkali or alkaline-earth atoms trapped in magnetic, magneto-optical, or dipole traps \cite{harter2014,cote2016,tomza2019,smialkowski2020}. 
Several cold atomic ion-atom combinations have already been experimentally investigated \cite{tomza2019,smialkowski2020}, including Ca atoms and Ba$^+$ ions, confined in a hybrid trap \cite{sullivan2012}. In addition, the interactions and chemical reactions between the neutral alkaline-earth-metal (A) and ionic alkaline-earth-metal (B) are also being studied for diatomic AB$^+$, and triatomic A$_2$B$^+$ systems \cite{smialkowski2020}. Due to the recent developments in the research field, there is a growing interest worldwide to investigate these systems, both experimentally and theoretically. With this motivation, we have carried out a thorough theoretical study of the long-range ($C_6$ and $C_9$) interactions of atom-ion between the alkaline-earth atoms and ions. 

In the present work, we have determined $C_6$ coefficient among  Mg, Ca, Sr, and Ba alkaline-earth atoms and with Mg$^+$, Ca$^+$, Sr$^+$, and Ba$^+$ alkaline-earth ions. Also, the $C_9$ coefficients for the interaction among three alkaline earth atoms, as well as two alkaline-earth atoms with an alkaline-earth ion are evaluated. Despite the importance, the dispersion coefficients of these systems are not explored experimentally or theoretically to date; therefore, we cannot compare our calculated data with the literature. To ascertain the reliability of our calculation procedure and the accuracy of the obtained results, we have calculated the dispersion  coefficients ($C_6$) for the homonuclear alkaline-earth atoms and compared with the previously available results \cite{li1973,allard2003}.

In order to obtain these dispersion coefficients, we have employed relativistic methods to calculate dynamic polarizabilities of atoms and ions. For this purpose, the oscillator strength of the leading transitions in alkaline-earth ions and atoms are calculated and compared with the National Institute of Standards and Technology (NIST)~\cite{NIST} as well as other \cite{risberg1955,edlen1956,sullivan1938,karlsson1999,meissner1938,risberg1965,risberg1968}  values. Results are also reported for ground state polarizability and their comparison is presented with other calculations \cite{lim2004,dere-seg} and measurements \cite{chang1983,nunkaew2009,snow2007,36,28,38,schwartz1974}, wherever available. Finally, $C_6$ coefficients for homonuclear dimers of alkaline-earth atoms are compared with other results \cite{li1973,allard2003}, whereas, for $C_6$ and $C_9$ values of the remaining combinations, we have used the approximate fitting models to verify our results.

The paper is organised as follows. In Section~\ref{theory}, we give a brief overview of the theoretical methodology employed in the present work. Our results are presented and discussed in Section~\ref{results}. Finally, conclusions are drawn in Section IV. We have used atomic units (a.u.) throughout the manuscript unless stated otherwise.    

\section{THEORETICAL CALCULATION}~\label{theory}

\subsection{Dispersion coefficients}
The long-range van der Waals interaction among three atomic species in ground states is given by
\begin{eqnarray}
V(\vec{R_{12}},\vec{R_{23}},\vec{R_{31}}) &=&  -\frac{C_6^{(12)}}{R^6_{12}}-\frac{C_8^{(12)}}{R^8_{12}}-\ldots-\frac{C_6^{(23)}}{R^6_{23}}-   	                                                                                   \frac{C_8^{(23)}}{R^8_{23}}\nonumber\\
                                                                      &-&\ldots-\frac{C_6^{(31)}}{R^6_{31}}-\frac{C_8^{(31)}}{R^8_{31}}-\ldots\nonumber\\
                                                                            &+&(1+3\cos\theta_1\cos\theta_2\cos\theta_3)\frac{C^{123}_9}{R^3_{12}R^3_{23}R^3_{31}},\label{pot}\nonumber\\
\end{eqnarray}
where $C_n^{(ij)}$ and $C_n^{(ijk)}$ parameters are the dispersion coefficients for two body and three body interaction respectively, with $i, j, k = 1, 2, 3$  and $R_{ij}$ is the inter atomic distance between the $i^{th}$ and $j^{th}$ atomic systems. The angle $\theta_{k}$ is defined as $\cos\theta_k=-\hat R_{ik}\cdot\hat R_{kj}$. The leading contributor to the potential is from the $C_6$  terms which are defined in terms of dipole polarizabilities $\alpha_{i}(\iota\omega)$ as
\begin{equation}
C_6^{(ij)}=\frac{3}{\pi}\int_0^{\infty}d\omega\alpha_{i}(\iota\omega)\alpha_{j}(\iota\omega). \label{ab-initioc6}
\end{equation}
The last term of the Eq.~\ref{pot} is the lowest order of inverse powers to the internuclear distances in the third-order correction to the ground-state energy and is a three-body interaction term which is given by~\cite{marin}
\begin{equation}
C_9^{(ijk)}=\frac{3}{\Pi}\int_0^{\infty}d\omega\alpha_{i}(\iota\omega)\alpha_{j}(\iota\omega)\alpha_{k}(\iota\omega).\label{ab-initioc9}
\end{equation}
Note that this term has a different sign  from those of the other pair interaction terms  in Eq.~\ref{pot} and is thus called the non additive part of the long-range interaction potential. It is multiplied by an angle dependent factor which is positive for max$(\theta_1, \theta_2, \theta_3)<117^{\circ}$ and negative for max$(\theta_1, \theta_2, \theta_3)>126^{\circ}$~\cite{marin}. In the present work, we determine $C_6$ and $C_9$ coefficients using the \textit{ab-initio} methods as given by Eq. 2 and 3. 
\\


In addition, we also calculate these coefficients using simple fitting formulae for comparison purposes. Since it is cumbersome to determine polarizabilities for a sufficiently large number of frequencies, therefore  instead of using the exact {\it ab initio} method, alternative fitting methods have been adopted to calculate the $C_6$  coefficients in the literature. Among these the Slater-Kirkwood formula~\cite{sk} is one of the mostly used method in which the  dispersion coefficients for the atom-ion system are approximated by
\begin{equation}
C_{6}^{(ij)}={\frac{3}{2}}{\frac{\alpha_{i}\alpha_{j}}{(\alpha_{i}/N_{i})^{1/2}+(\alpha_{j}/N_{j})^{1/2}}},
\end{equation}
where $N_{i}$ and $N_{j}$ are the effective number of electrons in i$^{th}$ and j$^{th}$ atomic systems respectively and both can be  determined using the following empirical formula which assumes that
the dominant contributions arise from the loosely bound electrons present in the outer shell of the systems
\begin{equation}
\label{number}\\
(N_{i})^{1/2}=\frac{4}{3}{C_6^{(ii)}}/{(\alpha_{i})^{3/2}}
\end{equation}
with the van der Waals coefficient $C_6^{ii}$ of the homo-nuclear dimer and static polarizability $\alpha_i$ of the atom $i$.
Substituting the above relation, we get 
\begin{equation}
C_{6}^{(ij)}=\frac{2C_6^{(ii)}C_6^{(jj)}}{(\frac{\alpha_{j}}{\alpha_{i}})C_6^{(ii)}+(\frac{\alpha_{i}}{\alpha_{j}})C_6^{(jj)}}.\label{slater}
\end{equation}
The above formula has been extensively tested by Kramer and Herschbach~\cite{kramer} , and found to give quite good estimate of the $C_6$ coefficients.
Similarly the non additive $C_9$ coefficients for the atomic three-dipole dispersion coefficient can be obtained  by using  an approximate fitting through the following expression derived by Midzuno and Kihara~\cite{midzuno}
\begin{equation}
C_9^{(iii)}=\frac{3}{4}\alpha_i(0)C_6^{(ii)}\label{fitc9-1}
\end{equation}
and for the general case of three different atoms, they obtain
\begin{equation}
C_9^{(ijk)}=\frac{2*Q^iQ^jQ^k(Q^i+Q^j+Q^k)}{(Q^i+Q^j)(Q^j+Q^k)(Q^k+Q^i)},\label{fitc9-2}
\end{equation}
where~\cite{tang1969} 
\begin{equation}
Q^i=\frac{\alpha_j(0)\alpha_k(0)}{\alpha_i(0)}C_6^{(ii)}.
\end{equation}
Nevertheless, both the above fitting formulae given by Eq. 6 and 8 are only valid and good for the qualitative description of the ion-atom dispersion coefficients, but it is imperative to use more accurate values of  polarizabilities for the precise description of the ion-atom dispersion coefficients. In the present work, we determine these quantities for the alkaline-earth ions  and  alkaline-earth atoms using the {\it ab initio} methods as given by Eqs.~\ref{ab-initioc6} and \ref{ab-initioc9} and further compare these coefficients with the values obtained using the Slater-Kirkwood formula  given by Eq.~\ref{slater} and with the approximations used by  Midzuno and Kihara~\cite{midzuno} (Eqs.~\ref{fitc9-1} and ~\ref{fitc9-2}), which we refer to as the fitted values in our discussion. Moreover, we also determine the oscillator strengths of the leading transition and static dipole polarizabilities of the ground states of the alkaline earth ions and atoms and compare them with the available experimental values and other precise calculations. 
\begin{table*}[h!]
\caption{Comparison of our calculated oscillator strengths ($f$) of the leading transitions with the previous available values. The numbers in parentheses represent powers of 10.}
\label{os}
\begin{ruledtabular}
\begin{tabular}{lccccc}
Ion & Upper level & Lower level & Term & $f_{\rm present}$ & $f_{\rm previous}$ \\
\hline
Mg$^+$ & $3p$ & $3s$ &  $^2S_{1/2}$$\rightarrow$$^2P_{1/2}$ & 0.303(-2)   &  0.303(-2)$^a$ \\
Mg$^+$ & $3p$ & $3s$ &  $^2S_{1/2}$$\rightarrow$$^2P_{3/2}$ & 0.609(-1)  & 0.608(-2)$^a$  \\
Ca$^+$ & $4p$ & $4s$ &  $^2S_{1/2}$$\rightarrow$$^2P_{1/2}$ &  0.321(-2)  & 0.330(-2)$^b$  \\
Ca$^+$ & $4p$ & $4s$ &  $^2S_{1/2}$$\rightarrow$$^2P_{3/2}$ &  0.648(-2)   &  0.682(-2)$^b$ \\
Sr$^+$ & $5p$ & $5s$ &  $^2S_{1/2}$$\rightarrow$$^2P_{1/2}$ &  0.344(-2)  & 0.341(-2)$^c$  \\
Sr$^+$ & $5p$ & $5s$ &  $^2S_{1/2}$$\rightarrow$$^2P_{3/2}$ &  0.710(-2)  &  0.703(-2)$^c$  \\
Ba$^+$ & $6p$ & $6s$ &  $^2S_{1/2}$$\rightarrow$$^2P_{1/2}$ &  0.342(-1)  &  0.348(-2)$^d$ \\
Ba$^+$ & $6p$ & $6s$ &  $^2S_{1/2}$$\rightarrow$$^2P_{3/2}$ & 0.739(-2)   & 0.690(-2)$^d$  \\
\hline
Atom & Upper level & Lower level & Term & $f_{\rm present}$ &   $f_{\rm previous}$ \\
\hline
Mg      &  $3s3p$      &   $3s^2$  & $^1S_0$$\rightarrow$$^3P_1$ & 0.309(-8) & 0.238(-7)$^e$ \\
Mg     &  $3s3p$      &   $3s^2$  & $^1S_0$$\rightarrow$$^1P_1$ & 0.175(1) &  0.180(1)$^f$\\
Ca      &  $4s4p$      &   $4s^2$  & $^1S_0$$\rightarrow$$^3P_1$ & 0.825(-5)  & 0.510(-6)$^g$ \\
Ca     &  $4s4p$      &   $4s^2$  & $^1S_0$$\rightarrow$$^1P_1$ & 0.175(1) & 0.175(1)$^g$ \\
Sr      &  $5s5p$      &   $5s^2$  & $^1S_0$$\rightarrow$$^3P_1$ & 0.351(-5) & 0.100(-4)$^c$ \\
Sr     &  $5s5p$      &   $5s^2$  & $^1S_0$$\rightarrow$$^1P_1$ & 0.199(1) & 0.192(1)$^c$\\
Ba      &  $6s6p$      &   $6s^2$  & $^1S_0$$\rightarrow$$ ^3P_1$ & 0.836(-4) &0.840(-4)$^d$  \\
Ba     &  $6s6p$      &   $6s^2$  & $^1S_0$$\rightarrow$$^1P_1$ & 0.162(1) & 0.164(1)$^d$
\end{tabular}
\end{ruledtabular}
$^a$Risberg\cite{risberg1955},$^b$Edlen \textit{et al.}\cite{edlen1956},$^c$Sullivan\cite{sullivan1938},$^d$Karlsson \textit{et al.}\cite{karlsson1999},$^e$Meissner\cite{meissner1938},$^f$Risberg\cite{risberg1965},$^g$Risberg\cite{risberg1968}
\end{table*}

\begin{table*}[h!]
\caption{Comparison of our calculated static dipole polarizabilities $\alpha^v(0)$ (in a.u.) for the considered alkaline-earth ions and alkaline-earth atoms in their ground states with the available measurements and the other theoretical calculations. }
\label{pol}
\begin{ruledtabular}
\begin{tabular}{lcccc}

Ion &  State & Present Calculations & Other Calculations  & Measurements   \\
\hline

Mg$^+$       &    3s  & 35.0&&  \\
Ca$^+$       &    4s    & 76.1&75.88~\cite{lim2004}& 75.3(4)~\cite{chang1983} \\
Sr$^+$       &    5s   & 91.7 & 91.10~\cite{lim2004},93.3(9)~\cite{49} & 86(11)~\cite{nunkaew2009} \\
Ba$^+$       &     6s  & 125.5 &123.07~\cite{lim2004}  & 123.88(5)~\cite{snow2007} \\
\hline
Atom &  State & Present Calculations & Other Calculations~\cite{dere-seg} & Measurements   \\
\hline
Mg     &    $3s^2$    &71.9 & 71.3(7)& 71.5(3.1)~\cite{36}\\
Ca      &    $4s^2$     &158.2 &157.1(1.3)& 168.7(13.5)~\cite{28} \\
Sr      &    $5s^2$    & 214.5 & 197.2(2)& 186(15)~\cite{38} \\
Ba      &     $6s^2$    & 276.2  & 273.5(2.0) & 268(22)~\cite{schwartz1974} \\
\end{tabular}
\end{ruledtabular}
\end{table*}

\subsection{Dipole Polarizability}
It follows that calculation of the dispersion coefficients require knowledge of the dipole $\alpha^{v}(\iota\omega)$ dynamic polarizabilities for the atom and ion in ground state $v$. 
Following~\cite{arora2012} $\alpha^{v}(\iota\omega)$ can be categorised into three parts,
\begin{equation}
\alpha^{v}(\iota\omega)=\alpha^{v}_c(\iota\omega)+\alpha^{v}_{val}(\iota\omega)+\alpha^{v}_{vc}(\iota\omega),
\end{equation}
with the notations $c$ and $val$ representing the contributions due to the core and valence effects, 
respectively whereas $\alpha^v_{vc}$  is the compensation term which accounts for the contribution from the excitation to the occupied valence 
shell that is forbidden by the Pauli exclusion principle. In sum-over-states approach the valence correlation contributions to the ground state can be estimated
using the expression
\begin{equation}
\alpha^{v}_{val} (\iota\omega) = \sum_{p \ne v} \frac{f_{vp}}{(E_v-E_p)^2+\omega^2},\label{pol-eq}
\end{equation}
where $f$ is the  oscillator strength from state $v$ to excited intermediate states $p$ and $E$s are the corresponding energies of the states. 
The oscillator strengths $f$ for the corresponding transitions can be deduced using the relation~\cite{pol-os}
\begin{equation}
f_{vp}=-\frac{303.756}{g_v\lambda}\times|\langle j_p||D||j_v\rangle|^2\label{em-os}.
\end{equation}
Here $\lambda$ is the transition wavelength expressed in \AA,  $g_v$ denotes the statistical weight and $\langle j_p||D||j_v\rangle$ represents the reduced dipole matrix element in a.u. 

It is required to calculate a sufficient number of atomic states so that ample oscillator strengths  can be evaluated to estimate contributions to the $\alpha^{v}_{val}$ values.  
We have evaluated as many transitions as possible for accurate calculation of the valence contribution by  either using all-order method and Multi-Configurational Dirac Fock (MCDF) approximation for ions and atoms respectively as discussed in next section. For better accuracy in the calculation of polarizabilities instead of using our calculated energy values, we use experimental energies from  NIST database, where  the  best  compiled  values  have  been  given \cite{NIST}.  
The core contributions $\alpha^{v}_c(\iota\omega)$ have been calculated using a random phase approximation (RPA) as described in Refs.~\cite{Blundell,singhy,kaurj}. 
The core-valence contributions $\alpha^{v}_{vc}(\iota\omega)$ are typically extremely small in magnitude and have been ignored in the present study.

\subsection{Evaluation of matrix element}~\label{matel}
The approach we use here to evaluate the dispersion coefficients is based on the use of accurately calculated dipole matrix elements for the two atomic states. 
The wavefunctions required for the evaluation of  matrix elements for alkaline-earth ions are obtained considering 
the singles and doubles excitation approximation in the all-order (SD) method as described in Refs. \cite{Blundell,theory}. In the SD 
method, the wave function of the state with the closed-core with a valence electron $v$ is represented 
as an expansion
\begin{eqnarray}
|\Psi_v \rangle_{\rm SD} &=& \left[1+ \sum_{ma}\rho_{ma} a^\dag_m a_a+ \frac{1}{2} \sum_{mnab} \rho_{mnab}a_m^\dag a_n^\dag a_b a_a\right. \nonumber\\
  &&\left. + \sum_{m \ne v} \rho_{mv} a^\dag_m a_v + \sum_{mna}\rho_{mnva} a_m^\dag a_n^\dag a_a a_v\right] |\Phi_v\rangle,\nonumber \\
&&  
  \label{sdmethod}
\end{eqnarray}
where $|\Phi_v\rangle$ is the Dirac Hartree Fock (DHF) wave function of the state. In the above expression, $a^\dag_i$ and $a_i$ are the creation and 
annihilation operators with the indices $\{m,n\}$ and $\{a,b\}$ designating the virtual and core orbitals of $|\Phi_v\rangle$, 
$\rho_{ma}$ and $\rho_{mv}$ are the corresponding single core and valence excitation coefficients, and $\rho_{mnab}$ and 
$\rho_{mnva}$ are the double core and valence excitation coefficients. To construct the single particle orbitals for the SD method, we have used total 70 B-spline functions 
with a cavity of radius $R=220$ a.u.

The wavefunctions required for the computation of matrix elements for alkaline-earth atoms in their initial and final states are obtained under MCDF approach using GRASP2k code~\cite{jonsson2013}. In this approximation an atomic state function (ASF) is considered as a linear combination of various configurational state functions (CSFs) which have same total angular momentum and parity, i.e.,
\begin{equation}
|\Psi_v\rangle_{\rm MCDF}=\sum_{n=1}^Na_n|\Phi_n\rangle 
\end{equation}
here, $n$ denotes the number of CSFs, $a_n$ is the mixing coefficient of the CSF $|\Phi_n\rangle$ in representation of the state $|\Psi_v\rangle$. The single particle orbital radial functions and expansion coefficients are obtained first by multi-configuration self-consistent field calculations using Dirac-Coulomb Hamiltonian. Further, relativistic configuration interaction calculations are performed including Breit and quantum electrodynamic corrections. In order to increase the accuracy of the ASF we consider maximum number of CSFs in the linear contribution and finally, retain only those which have value of mixing coefficient greater than  10$^{-3}$. It is important to mention here that in the present work we have used all order method for ions as this method is suitable for the monovalent system and give accurate result for them as compared to Multi-configuration Dirac-Fock (MCDF) method. Since for the two valence electron alkaline earth atoms it is difficult to apply all order method and also at present we do not have a suitable code developed yet, we have used the MCDF approach and utilized available GRAPS2k \cite{jonsson2013} which gives equally accurate results for divalent systems. 

After obtaining wave functions for the aforementioned ions and atoms, we determine the dipole allowed (E1) matrix element for a transition. 
It is relatively straight forward to make use of the generated list of matrix elements to evaluate oscillator strengths and  dipole polarizabilities which are further used to calculate the dispersion coefficients as described in the theory earlier.

\begin{table*}[h!]
\caption{ The dispersion coefficients ($C_{6}$) for the interaction between two homonuclear alkaline-earth atoms in ground state.
\label{c6-atom-atom1}}
\begin{ruledtabular}
\begin{tabular}{lccc}
             & Present Calculation & Other Calculations & Measurements  \\ 
\hline
Mg-Mg &  636 & 612.6$^a$,627(12)$^b$,648$^c$ & 683(35)$^d$ \\
Ca-Ca  & 2138 & 2022$^a$,2221(15)$^b$,2002$^c$ & 2080(7)$^e$ \\
Sr-Sr    & 3654 & 2890$^a$,3170(196)$^b$,2849$^c$ & - \\
Ba-Ba  &  5324 & 5160(74)$^b$,4479$^c$ & -       
\end{tabular}
\end{ruledtabular}
$^a$Mitroy \textit{et al.}\cite{mitroy2003}, $^b$Porsev  \textit{et al.}\cite{porsev2006}, $^c$Patil\cite{patil2000}, $^d$Kwong \textit{et al.}\cite{li1973}, $^e$Allard \textit{et al.}\cite{allard2003}.
\end{table*}

\begin{table*}[h!]
\caption{ The dispersion coefficients ($C_{6}$) for the interaction between two heteronuclear alkaline-earth atoms and their comparison with the fitting result obtained by Eq.~\ref{slater}.
\label{c6-atom-atom2}}
\begin{ruledtabular}
\begin{tabular}{lcc}
             & Present Calculation & Fitted\\ 
\hline
Mg-Ca & 1150 &  1160  \\
Mg-Sr &  1494 & 1380  \\
Mg-Ba & 1784 &  1727  \\
Ca-Sr &  2793&  2652 \\
Ca-Ba &  3360&3367   \\
Sr-Ba & 4403 &  4032 \\  
\end{tabular}
\end{ruledtabular}
\end{table*}

\begin{table*}[h!]
\caption{The dispersion coefficients ($C_{6}$) for the alkaline-earth ions interacting with the alkaline-earth atoms and their comparison with the fitting value obtained using Eq.~\ref{slater} \label{c6-ion-atom}.}
\begin{ruledtabular}
\begin{tabular}{lccccccccc}
                     & \multicolumn{2}{c}{Mg}   & \multicolumn{2}{c}{Ca}   & \multicolumn{2}{c}{Sr}  & \multicolumn{2}{c}{Ba}       \\
                    & Present Cal. & Fitted & Present Cal. & Fitted& Present Cal. & Fitted & Present Cal. & Fitted\\
\hline
& &	\\
Mg$^+$ & 312&   310    & 564  &574 &  733& 682&  876	& 853 \\
\\
Ca$^+$	 & 587 & 584       & 1082 &  1105   &1413 &     1317     &  1698& 	1664\\
 \\
Sr$^+$	 & 706 &   702      &1302  &   1335   & 1699 &   1593      & 2044& 	2016\\
 \\
Ba$^+$& 930&    930    & 1719&   1770  &2247 &   2112      &	2708& 2673 \\ \\
\end{tabular}
\end{ruledtabular}
\end{table*}

\section{Results}~\label{results}

\subsection{Oscillator strengths}
We have calculated the first few oscillator strengths having dominant contributions to the dipole polarizability as given by Eq.~\ref{pol-eq}. In Table~\ref{os}, a comparison is presented  between present calculations with the previously reported other results \cite{risberg1955,edlen1956,sullivan1938,karlsson1999,meissner1938,risberg1965,risberg1968}  which  are also available at the NIST database \cite{NIST}.  We find excellent agreement between the two results for ions. However, for atoms, the calculated oscillator strengths agree well with the previous data for $^1S_0 \rightarrow ^1P_1$ transition, while for $^1S_0 \rightarrow ^3P_1$ transitions, a slight discrepancy of the order of $\sim $ 10$^{-6}$ can be observed. This difference can be conveniently ignored as the small value of oscillator strength will have an insignificant contribution to the determination of polarizability. Thus, our results for oscillator strengths are precise enough to predict the reliable values of the polarizability of the considered ionic and atomic systems.

\subsection{Static dipole polarizabilities}
 The calculated oscillator strengths are used to determine the static dipole polarizabilities of the concerned alkaline-earth ions and atoms. As described previously in theoretical method we use a sum-over-states approach to calculate the polarizability in this work. It should however be noted that for the ground state polarizability only first few low lying transitions contribute the most. Please note that the error introduced due to truncation of the sumation was not more than 1.5\% for ions and even smaller for the atoms.

\textit{Alkaline-earth ions:} \\
Our results for static dipole polarizability of alkaline-earth ions are presented and compared in Table~\ref{pol} with other theoretical calculations of Lim and Schwerdtfeger \cite{lim2004} as well as different precise measurements \cite{chang1983,nunkaew2009,snow2007}.  One can readily see that present results are in good agreement with the theoretical results \cite{lim2004} that are obtained using the relativistic coupled-cluster method in the finite field gradient technique along with the optimized Gaussian-type basis set. However, in our calculation, we have used a sum-over-sates approach which permits us to use accurate experimental energy data wherever available. 
This justifies the good agreement of the present calculations with the measurements. 
 For example, the static dipole polarizability value of Ca$^+$ ion, measured by Chang \textit{et al.} \cite{chang1983}, is in very close agreement with our evaluated value, whereas, for other ions, i.e., Sr$^+$ \cite{nunkaew2009} and Ba$^+$ \cite{snow2007}, the agreement is within 2\%. Unfortunately, for Mg$^+$ ion we did not find any data in the literature to compare with our result. However, based on our results of Ca$^+$,  Sr$^+$ and Ba$^+$ ions, we believe our value for Mg$^+$ ion should also be reliable.

\textit{Alkaline-earth atoms:}\\
The present results of static dipole polarizability of alkaline-earth atoms are given in Table~\ref{pol} where these are compared with the previously reported calculation of Porsev and Derevianko \cite{porsev2006} and other experimental results \cite{36,28,38,schwartz1974}.  Porsev and Derevianko calculated their static dipole polarizabilities by using available reported experimental energies and theoretically calculated matrix elements. A close agreement between our results for Mg, Ca, and Ba atoms with theoretical calculations \cite{porsev2006}  and other experimental measurements \cite{36,28,38,schwartz1974} can be seen from Table~\ref{pol}. However, for Sr atom, we observe that our value of static dipole polarizability is somewhat higher as compared to the available calculation \cite{porsev2006} and the experimental value \cite{38}. This could be due to our slightly higher value of oscillator strength as compared to that of the NIST \cite{NIST} value for the leading transition $^1S_0 \rightarrow ^1P_1$.


\begin{table*}[h!]
\caption{ The dispersion coefficients ($C_{9}$) for the interaction among three alkaline-earth atoms. Fitted values are obtained from Eqs.~\ref{fitc9-1}. \label{c9-atom-atom-atom} }
\begin{ruledtabular}
\begin{tabular}{lccccc}
                   &  &  Mg-Mg   & Ca-Ca    & Sr-Sr   & Ba-Ba       \\
\hline
& &	\\
Mg&Present Calculation & 34024 & 125721 &  217665& 322323	\\
                          & Fitted        &     33482       &        131571       &         190085    &     319626       \\
\\
Ca&Present Calculation	& 64768    & 248117 & 432498 & 647488 	\\
                              & Fitted        &    65779         &      267852    &   388883      &        663115      \\
\\
Sr&Present Calculation	& 84848&  327456 & 571661 & 858017	\\
                             & Fitted        &      78724        &       322681     &     468843   &        801218         \\
\\
Ba&Present Calculation & 102290 & 399639 & 699526 &	1054803   \\
                              & Fitted       &      100814       &    420111        &      611823      &     1052640\\
\\
\end{tabular}
\end{ruledtabular}
\end{table*}

\begin{table*}[h!]
\caption{ The dispersion coefficients ($C_{9}$) for the interaction between two alkaline-earth atoms and alkaline-earth ions. The fitted values are obtained using Eq.~\ref{fitc9-2}.\label{c9-atom-atom-ion} }
\begin{ruledtabular}
\begin{tabular}{lccccc}
                    & &  Mg-Mg   & Ca-Ca    & Sr-Sr   & Ba-Ba       \\
\hline
& &	\\
Mg$^+$&Present Calculation &  16634 &  61381 &  106248 &	157285 \\
                                     & Fitted &     16537    &   64896      &       93742       &        157547       \\
\\
Ca$^+$	&Present Calculation&  32308 & 122618 & 213393 & 318658 	\\
                                    & Fitted &      32420     &130467         &       189086     &     320800          \\
\\
Sr$^+$&Present Calculation&	38619 &  146658&  255303 & 381479  	\\
                                    & Fitted &     39314       &     159066        &     230705         &      392236        \\
\\
Ba$^+$&Present Calculation	& 50859 &  194466&  339068&508069	  \\
                                    & Fitted &       52133      &      210998     &     306040       &     520380        \\
\\
\end{tabular}
\end{ruledtabular}
\end{table*}
\subsection{$C_6$ coefficient}
Using our calculations for dipole polarizabilities, we further calculate the $C_6$ dispersion coefficients for the combination of interaction between two homonuclear alkaline-earth atoms, two heteronuclear alakline-earth atoms as well as alkaline-earth atom with their ions in their ground state. We have taken the calculation of C6 for homonuclear alkaline-earth atom dimers simply for comparison pourposes in order to just check the reliability of our calculations and to see how these match with the other available results. These coefficients are presented in Table~\ref{c6-atom-atom1}, where we have also compared our calculations with the available previous theoretical \cite{mitroy2003,porsev2006,patil2000} and experimental \cite{li1973,allard2003} results. We find that our calculated value of $C_6$ for Ca-Ca dimer shows excellent agreement with the experimental result while it deviates from measurements by $\sim $ 7\% for Mg-Mg dimer. Also, one can see from Table~\ref{c6-atom-atom1}, that there is significant disagreement among the reported theoretical values from  Mitroy \textit{et al.} \cite{mitroy2003}, Porsev \textit{et al.} \cite{porsev2006}, and Patil \textit{et al.} \cite{patil2000}, for all homonuclear alkaline-earth atom dimers. This difference among these results can be attributed to different methodologies adopted in the theoretical approaches. We also observe that with the increase in the atomic size of homonuclear dimer the $C_6$ dispersion coefficients also increase. The same trend can also be seen in the other reported theoretical as well as experimental results.

Further, we have calculated $C_6$ dispersion coefficients for the interaction between the two heteronuclear alkaline-earth atoms in their ground state and these are shown in the Table~\ref{c6-atom-atom2}. There are no other results reported for these $C_6$ coefficients with which we can compare our results. However, for the sake of comparison, we have also obtained $C_6$ using Slater Kirkwood formula given by Eq.~\ref{slater} using experimental values of static dipole polarizabilities of each atom as well as  $C_6$ coefficients as reported by Porsev \textit{et al.} \cite{porsev2006} for homonuclear alkaline-earth atom dimers. These calculated values are also presented for comparison in Table~\ref{c6-atom-atom2} and referred as fitted calculations. Results determined using this approach show a close agreement with the present $C_6$ values, except for the dimer which has a combination with Sr atom, i.e., Mg-Sr, Ca-Sr and Sr-Ba dimers. This could be due to our calculated dipole polarizability of Sr atom being relatively higher than the experimental value as can be seen from Table II. An agreement between results from both the methods clearly confirms the validity of our method of calculation of $C_6$ dispersion coefficients for heteronuclear alkaline-earth atoms. Similar to the previous case, we again find that the value of  $C_6$ increases with the increasing size of interacting atoms.

The calculated $C_6$ dispersion coefficients for the alkaline-earth ions interacting with the alkaline-earth atoms are shown in Table~\ref{c6-ion-atom}. In this case, also there are no other previously reported theoretical or experimental results to compare with our results. Consequently, we performed calculations using fitting approach using eq.~\ref{slater} and compared these in the Table~\ref{c6-ion-atom}. We observe good agreement between present results with our fitting calculations for all the dimers of alkaline-earth ions with the alkaline-earth atoms except where neutral Sr atom is involved. The possible reason can again be due to the relatively higher value of our calculated dipole polarizability. The close agreement between these two approaches justifies the accuracy of our method of obtaining $C_6$ coefficients.

\subsection{$C_9$ coefficients}
After calculating the value of $C_6$ for dimers, we have determined the $C_9$ dispersion coefficients for the interaction between three alkaline-earth atoms. The calculated results are shown in Table~\ref{c9-atom-atom-atom}. Due to non-availability of other results for comparisons, we have again evaluated $C_9$  using fitting equations, i.e., Eqs.~\ref{fitc9-1} and \ref{fitc9-2}. In order to solve these equations, we have used $C_6$ coefficients for homonuclear alkaline-earth atom dimers by Porsev \textit{et al.} \cite{porsev2006} and used experimental static dipole polarizability from Peter \& Jeffrey \cite{schwerdtfeger2019}. One can observe from Table~\ref{c9-atom-atom-atom}, that both sets of values for $C_9$ dispersion coefficients of the concerned systems show close agreement. The maximum difference between two results can be seen for the systems which have Sr atom in their combinations for a reason as stated earlier.

Finally, we have calculated the values of $C_9$, as given in Table~\ref{c9-atom-atom-ion}, for the interaction between two alkaline-earth atoms and alkaline-earth ions. In addition to a calculation by our method, we have determined these coefficients with fitting equations, i.e., Eqs.~\ref{fitc9-1} and \ref{fitc9-2}.  To solve these equations experimental values \cite{schwerdtfeger2019} of the static dipole polarizability  Ca$^+$, Sr$^+$, and Ba$^+$ are used. In case of Mg$^+$ our calculated result for dipole polarizability is utilized due to unavailability of the corresponding measurements. From this table, we find close agreement between both sets of the calculations, excluding the systems with Sr atom. This is consistent with our other results for $C_6$ and $C_9$. The $C_9$ coefficients also show similar trends as $C_6$ with respect to the size of the interacting atomic systems.

\section{Conclusion}
In the present work, we have studied the nature of the interaction coefficients for the alkaline-earth atoms among themselves and with the alkaline-earth ions and obtained the values of the two-body $C_6$ and three-body $C_9$ dispersion coefficients. To determine these coefficients, oscillator strengths for leading transitions and static dipole polarizability of atoms and ions in their ground states are determined using relativistic methods. These results  are compared with corresponding data from other theoretical calculations, measurements and NIST database and an overall, good agreement are found. Apart from  $C_6$ values for alkaline-earth homonuclear dimers, the  $C_6$ and  $C_9$ coefficients are reported here for the first time. Therefore, we have performed fitting calculations, for the sake of comparison, using measured values of dipole polarizability and other available parameters. A good agreement between values from our and fitting methods indicates the reliability of our calculations for  $C_6$ and  $C_9$ dispersion coefficients. We hope our results will induce more theoretical and experimental studies in this direction and help to make progress in quantum information processing, quantifying molecular potentials for ultracold collision investigation and designing better atomic clocks.

\section*{Acknowledgements}

The work of B.A. is supported by DST-SERB Grant No. EMR/2016/001228. One of the authors (N.S.) is thankful to the Ministry of Human
Resources and Development (MHRD), Govt. of India for fellowship. RS and LS are thankful to the SERB-DST and CSIR,
New Delhi, Govt. of India for the sanction of research projects.

\bibliographystyle{apsrev4-1}

\begin{thebibliography}{94}%
\makeatletter
\providecommand \@ifxundefined [1]{%
 \@ifx{#1\undefined}
}%
\providecommand \@ifnum [1]{%
 \ifnum #1\expandafter \@firstoftwo
 \else \expandafter \@secondoftwo
 \fi
}%
\providecommand \@ifx [1]{%
 \ifx #1\expandafter \@firstoftwo
 \else \expandafter \@secondoftwo
 \fi
}%
\providecommand \natexlab [1]{#1}%
\providecommand \enquote  [1]{``#1''}%
\providecommand \bibnamefont  [1]{#1}%
\providecommand \bibfnamefont [1]{#1}%
\providecommand \citenamefont [1]{#1}%
\providecommand \href@noop [0]{\@secondoftwo}%
\providecommand \href [0]{\begingroup \@sanitize@url \@href}%
\providecommand \@href[1]{\@@startlink{#1}\@@href}%
\providecommand \@@href[1]{\endgroup#1\@@endlink}%
\providecommand \@sanitize@url [0]{\catcode `\\12\catcode `\$12\catcode
  `\&12\catcode `\#12\catcode `\^12\catcode `\_12\catcode `\%12\relax}%
\providecommand \@@startlink[1]{}%
\providecommand \@@endlink[0]{}%
\providecommand \url  [0]{\begingroup\@sanitize@url \@url }%
\providecommand \@url [1]{\endgroup\@href {#1}{\urlprefix }}%
\providecommand \urlprefix  [0]{URL }%
\providecommand \Eprint [0]{\href }%
\providecommand \doibase [0]{http://dx.doi.org/}%
\providecommand \selectlanguage [0]{\@gobble}%
\providecommand \bibinfo  [0]{\@secondoftwo}%
\providecommand \bibfield  [0]{\@secondoftwo}%
\providecommand \translation [1]{[#1]}%
\providecommand \BibitemOpen [0]{}%
\providecommand \bibitemStop [0]{}%
\providecommand \bibitemNoStop [0]{.\EOS\space}%
\providecommand \EOS [0]{\spacefactor3000\relax}%
\providecommand \BibitemShut  [1]{\csname bibitem#1\endcsname}%
\let\auto@bib@innerbib\@empty
\bibitem [{\citenamefont {Smith}\ \emph {et~al.}(2005)\citenamefont {Smith},
  \citenamefont {Makarov},\ and\ \citenamefont {Lin}}]{smith2005}%
  \BibitemOpen
  \bibfield  {author} {\bibinfo {author} {\bibfnamefont {W.~W.}\ \bibnamefont
  {Smith}}, \bibinfo {author} {\bibfnamefont {O.~P.}\ \bibnamefont {Makarov}},
  \ and\ \bibinfo {author} {\bibfnamefont {J.}~\bibnamefont {Lin}},\
  }\href@noop {} {\bibfield  {journal} {\bibinfo  {journal} {Journal of Modern
  Optics}\ }\textbf {\bibinfo {volume} {52}},\ \bibinfo {pages} {2253}
  (\bibinfo {year} {2005})}\BibitemShut {NoStop}%
\bibitem [{\citenamefont {Zipkes}\ \emph
  {et~al.}(2010{\natexlab{a}})\citenamefont {Zipkes}, \citenamefont {Palzer},
  \citenamefont {Sias},\ and\ \citenamefont {K{\"o}hl}}]{zipkes2010}%
  \BibitemOpen
  \bibfield  {author} {\bibinfo {author} {\bibfnamefont {C.}~\bibnamefont
  {Zipkes}}, \bibinfo {author} {\bibfnamefont {S.}~\bibnamefont {Palzer}},
  \bibinfo {author} {\bibfnamefont {C.}~\bibnamefont {Sias}}, \ and\ \bibinfo
  {author} {\bibfnamefont {M.}~\bibnamefont {K{\"o}hl}},\ }\href@noop {}
  {\bibfield  {journal} {\bibinfo  {journal} {Nature}\ }\textbf {\bibinfo
  {volume} {464}},\ \bibinfo {pages} {388} (\bibinfo {year}
  {2010}{\natexlab{a}})}\BibitemShut {NoStop}%
\bibitem [{\citenamefont {Aymar}\ \emph {et~al.}(2011)\citenamefont {Aymar},
  \citenamefont {Gu{\'e}rout},\ and\ \citenamefont {Dulieu}}]{aymar2011}%
  \BibitemOpen
  \bibfield  {author} {\bibinfo {author} {\bibfnamefont {M.}~\bibnamefont
  {Aymar}}, \bibinfo {author} {\bibfnamefont {R.}~\bibnamefont {Gu{\'e}rout}},
  \ and\ \bibinfo {author} {\bibfnamefont {O.}~\bibnamefont {Dulieu}},\
  }\href@noop {} {\bibfield  {journal} {\bibinfo  {journal} {The Journal of
  chemical physics}\ }\textbf {\bibinfo {volume} {135}},\ \bibinfo {pages}
  {064305} (\bibinfo {year} {2011})}\BibitemShut {NoStop}%
\bibitem [{\citenamefont {Gacesa}\ \emph {et~al.}(2016)\citenamefont {Gacesa},
  \citenamefont {Montgomery~Jr}, \citenamefont {Michels},\ and\ \citenamefont
  {C{\^o}t{\'e}}}]{gacesa2016}%
  \BibitemOpen
  \bibfield  {author} {\bibinfo {author} {\bibfnamefont {M.}~\bibnamefont
  {Gacesa}}, \bibinfo {author} {\bibfnamefont {J.~A.}\ \bibnamefont
  {Montgomery~Jr}}, \bibinfo {author} {\bibfnamefont {H.~H.}\ \bibnamefont
  {Michels}}, \ and\ \bibinfo {author} {\bibfnamefont {R.}~\bibnamefont
  {C{\^o}t{\'e}}},\ }\href@noop {} {\bibfield  {journal} {\bibinfo  {journal}
  {Physical Review A}\ }\textbf {\bibinfo {volume} {94}},\ \bibinfo {pages}
  {013407} (\bibinfo {year} {2016})}\BibitemShut {NoStop}%
\bibitem [{\citenamefont {Roberts}\ \emph {et~al.}(1998)\citenamefont
  {Roberts}, \citenamefont {Claussen}, \citenamefont {Burke~Jr}, \citenamefont
  {Greene}, \citenamefont {Cornell},\ and\ \citenamefont
  {Wieman}}]{roberts1998}%
  \BibitemOpen
  \bibfield  {author} {\bibinfo {author} {\bibfnamefont {J.}~\bibnamefont
  {Roberts}}, \bibinfo {author} {\bibfnamefont {N.}~\bibnamefont {Claussen}},
  \bibinfo {author} {\bibfnamefont {J.~P.}\ \bibnamefont {Burke~Jr}}, \bibinfo
  {author} {\bibfnamefont {C.~H.}\ \bibnamefont {Greene}}, \bibinfo {author}
  {\bibfnamefont {E.~A.}\ \bibnamefont {Cornell}}, \ and\ \bibinfo {author}
  {\bibfnamefont {C.}~\bibnamefont {Wieman}},\ }\href@noop {} {\bibfield
  {journal} {\bibinfo  {journal} {Physical Review Letters}\ }\textbf {\bibinfo
  {volume} {81}},\ \bibinfo {pages} {5109} (\bibinfo {year}
  {1998})}\BibitemShut {NoStop}%
\bibitem [{\citenamefont {Amiot}\ and\ \citenamefont
  {Verges}(2000)}]{amiot2000}%
  \BibitemOpen
  \bibfield  {author} {\bibinfo {author} {\bibfnamefont {C.}~\bibnamefont
  {Amiot}}\ and\ \bibinfo {author} {\bibfnamefont {J.}~\bibnamefont {Verges}},\
  }\href@noop {} {\bibfield  {journal} {\bibinfo  {journal} {The Journal of
  Chemical Physics}\ }\textbf {\bibinfo {volume} {112}},\ \bibinfo {pages}
  {7068} (\bibinfo {year} {2000})}\BibitemShut {NoStop}%
\bibitem [{\citenamefont {Leo}\ \emph {et~al.}(2000)\citenamefont {Leo},
  \citenamefont {Williams},\ and\ \citenamefont {Julienne}}]{leo2000}%
  \BibitemOpen
  \bibfield  {author} {\bibinfo {author} {\bibfnamefont {P.~J.}\ \bibnamefont
  {Leo}}, \bibinfo {author} {\bibfnamefont {C.~J.}\ \bibnamefont {Williams}}, \
  and\ \bibinfo {author} {\bibfnamefont {P.~S.}\ \bibnamefont {Julienne}},\
  }\href@noop {} {\bibfield  {journal} {\bibinfo  {journal} {Physical review
  letters}\ }\textbf {\bibinfo {volume} {85}},\ \bibinfo {pages} {2721}
  (\bibinfo {year} {2000})}\BibitemShut {NoStop}%
\bibitem [{\citenamefont {Leanhardt}\ \emph {et~al.}(2003)\citenamefont
  {Leanhardt}, \citenamefont {Shin}, \citenamefont {Chikkatur}, \citenamefont
  {Kielpinski}, \citenamefont {Ketterle},\ and\ \citenamefont
  {Pritchard}}]{leanhardt2003}%
  \BibitemOpen
  \bibfield  {author} {\bibinfo {author} {\bibfnamefont {A.}~\bibnamefont
  {Leanhardt}}, \bibinfo {author} {\bibfnamefont {Y.}~\bibnamefont {Shin}},
  \bibinfo {author} {\bibfnamefont {A.}~\bibnamefont {Chikkatur}}, \bibinfo
  {author} {\bibfnamefont {D.}~\bibnamefont {Kielpinski}}, \bibinfo {author}
  {\bibfnamefont {W.}~\bibnamefont {Ketterle}}, \ and\ \bibinfo {author}
  {\bibfnamefont {D.}~\bibnamefont {Pritchard}},\ }\href@noop {} {\bibfield
  {journal} {\bibinfo  {journal} {Physical review letters}\ }\textbf {\bibinfo
  {volume} {90}},\ \bibinfo {pages} {100404} (\bibinfo {year}
  {2003})}\BibitemShut {NoStop}%
\bibitem [{\citenamefont {Yuju}\ \emph {et~al.}(2004)\citenamefont {Yuju},
  \citenamefont {Teper}, \citenamefont {Cheng},\ and\ \citenamefont
  {Vuletic}}]{yuju2004}%
  \BibitemOpen
  \bibfield  {author} {\bibinfo {author} {\bibfnamefont {L.}~\bibnamefont
  {Yuju}}, \bibinfo {author} {\bibfnamefont {I.}~\bibnamefont {Teper}},
  \bibinfo {author} {\bibfnamefont {C.}~\bibnamefont {Cheng}}, \ and\ \bibinfo
  {author} {\bibfnamefont {V.}~\bibnamefont {Vuletic}},\ }\href@noop {}
  {\bibfield  {journal} {\bibinfo  {journal} {Phys. Rev. Lett.}\ }\textbf
  {\bibinfo {volume} {92}},\ \bibinfo {pages} {050404} (\bibinfo {year}
  {2004})}\BibitemShut {NoStop}%
\bibitem [{\citenamefont {Schmid}\ \emph {et~al.}(2010)\citenamefont {Schmid},
  \citenamefont {H{\"a}rter},\ and\ \citenamefont {Denschlag}}]{schmid2010}%
  \BibitemOpen
  \bibfield  {author} {\bibinfo {author} {\bibfnamefont {S.}~\bibnamefont
  {Schmid}}, \bibinfo {author} {\bibfnamefont {A.}~\bibnamefont {H{\"a}rter}},
  \ and\ \bibinfo {author} {\bibfnamefont {J.~H.}\ \bibnamefont {Denschlag}},\
  }\href@noop {} {\bibfield  {journal} {\bibinfo  {journal} {Physical review
  letters}\ }\textbf {\bibinfo {volume} {105}},\ \bibinfo {pages} {133202}
  (\bibinfo {year} {2010})}\BibitemShut {NoStop}%
\bibitem [{\citenamefont {Bodo}\ \emph {et~al.}(2008)\citenamefont {Bodo},
  \citenamefont {Zhang},\ and\ \citenamefont {Dalgarno}}]{bodo2008}%
  \BibitemOpen
  \bibfield  {author} {\bibinfo {author} {\bibfnamefont {E.}~\bibnamefont
  {Bodo}}, \bibinfo {author} {\bibfnamefont {P.}~\bibnamefont {Zhang}}, \ and\
  \bibinfo {author} {\bibfnamefont {A.}~\bibnamefont {Dalgarno}},\ }\href@noop
  {} {\bibfield  {journal} {\bibinfo  {journal} {New Journal of Physics}\
  }\textbf {\bibinfo {volume} {10}},\ \bibinfo {pages} {033024} (\bibinfo
  {year} {2008})}\BibitemShut {NoStop}%
\bibitem [{\citenamefont {Zhang}\ \emph {et~al.}(2009)\citenamefont {Zhang},
  \citenamefont {Bodo},\ and\ \citenamefont {Dalgarno}}]{zhang2009}%
  \BibitemOpen
  \bibfield  {author} {\bibinfo {author} {\bibfnamefont {P.}~\bibnamefont
  {Zhang}}, \bibinfo {author} {\bibfnamefont {E.}~\bibnamefont {Bodo}}, \ and\
  \bibinfo {author} {\bibfnamefont {A.}~\bibnamefont {Dalgarno}},\ }\href@noop
  {} {\bibfield  {journal} {\bibinfo  {journal} {The Journal of Physical
  Chemistry A}\ }\textbf {\bibinfo {volume} {113}},\ \bibinfo {pages} {15085}
  (\bibinfo {year} {2009})}\BibitemShut {NoStop}%
\bibitem [{\citenamefont {Peng~Zhang}\ and\ \citenamefont
  {Cote}(2009)}]{zhang20091}%
  \BibitemOpen
  \bibfield  {author} {\bibinfo {author} {\bibfnamefont {A.~D.}\ \bibnamefont
  {Peng~Zhang}}\ and\ \bibinfo {author} {\bibfnamefont {R.}~\bibnamefont
  {Cote}},\ }\href@noop {} {\bibfield  {journal} {\bibinfo  {journal} {Physical
  Review A}\ }\textbf {\bibinfo {volume} {80}},\ \bibinfo {pages} {030703}
  (\bibinfo {year} {2009})}\BibitemShut {NoStop}%
\bibitem [{\citenamefont {Idziaszek}\ \emph {et~al.}(2009)\citenamefont
  {Idziaszek}, \citenamefont {Calarco}, \citenamefont {Julienne},\ and\
  \citenamefont {Simoni}}]{idziaszek2009}%
  \BibitemOpen
  \bibfield  {author} {\bibinfo {author} {\bibfnamefont {Z.}~\bibnamefont
  {Idziaszek}}, \bibinfo {author} {\bibfnamefont {T.}~\bibnamefont {Calarco}},
  \bibinfo {author} {\bibfnamefont {P.~S.}\ \bibnamefont {Julienne}}, \ and\
  \bibinfo {author} {\bibfnamefont {A.}~\bibnamefont {Simoni}},\ }\href@noop {}
  {\bibfield  {journal} {\bibinfo  {journal} {Physical Review A}\ }\textbf
  {\bibinfo {volume} {79}},\ \bibinfo {pages} {010702} (\bibinfo {year}
  {2009})}\BibitemShut {NoStop}%
\bibitem [{\citenamefont {Gao}(2010)}]{gao2010}%
  \BibitemOpen
  \bibfield  {author} {\bibinfo {author} {\bibfnamefont {B.}~\bibnamefont
  {Gao}},\ }\href@noop {} {\bibfield  {journal} {\bibinfo  {journal} {Physical
  review letters}\ }\textbf {\bibinfo {volume} {104}},\ \bibinfo {pages}
  {213201} (\bibinfo {year} {2010})}\BibitemShut {NoStop}%
\bibitem [{\citenamefont {Cote}(2000)}]{cote2000}%
  \BibitemOpen
  \bibfield  {author} {\bibinfo {author} {\bibfnamefont {R.}~\bibnamefont
  {Cote}},\ }\href@noop {} {\bibfield  {journal} {\bibinfo  {journal} {Physical
  review letters}\ }\textbf {\bibinfo {volume} {85}},\ \bibinfo {pages} {5316}
  (\bibinfo {year} {2000})}\BibitemShut {NoStop}%
\bibitem [{\citenamefont {C{\^o}t{\'e}}\ and\ \citenamefont
  {Dalgarno}(2000)}]{cote20001}%
  \BibitemOpen
  \bibfield  {author} {\bibinfo {author} {\bibfnamefont {R.}~\bibnamefont
  {C{\^o}t{\'e}}}\ and\ \bibinfo {author} {\bibfnamefont {A.}~\bibnamefont
  {Dalgarno}},\ }\href@noop {} {\bibfield  {journal} {\bibinfo  {journal}
  {Physical Review A}\ }\textbf {\bibinfo {volume} {62}},\ \bibinfo {pages}
  {012709} (\bibinfo {year} {2000})}\BibitemShut {NoStop}%
\bibitem [{\citenamefont {Makarov}\ \emph {et~al.}(2003)\citenamefont
  {Makarov}, \citenamefont {C{\^o}t{\'e}}, \citenamefont {Michels},\ and\
  \citenamefont {Smith}}]{makarov2003}%
  \BibitemOpen
  \bibfield  {author} {\bibinfo {author} {\bibfnamefont {O.~P.}\ \bibnamefont
  {Makarov}}, \bibinfo {author} {\bibfnamefont {R.}~\bibnamefont
  {C{\^o}t{\'e}}}, \bibinfo {author} {\bibfnamefont {H.}~\bibnamefont
  {Michels}}, \ and\ \bibinfo {author} {\bibfnamefont {W.}~\bibnamefont
  {Smith}},\ }\href@noop {} {\bibfield  {journal} {\bibinfo  {journal}
  {Physical Review A}\ }\textbf {\bibinfo {volume} {67}},\ \bibinfo {pages}
  {042705} (\bibinfo {year} {2003})}\BibitemShut {NoStop}%
\bibitem [{\citenamefont {Doerk}\ \emph {et~al.}(2010)\citenamefont {Doerk},
  \citenamefont {Idziaszek},\ and\ \citenamefont {Calarco}}]{doerk2010}%
  \BibitemOpen
  \bibfield  {author} {\bibinfo {author} {\bibfnamefont {H.}~\bibnamefont
  {Doerk}}, \bibinfo {author} {\bibfnamefont {Z.}~\bibnamefont {Idziaszek}}, \
  and\ \bibinfo {author} {\bibfnamefont {T.}~\bibnamefont {Calarco}},\
  }\href@noop {} {\bibfield  {journal} {\bibinfo  {journal} {Physical Review
  A}\ }\textbf {\bibinfo {volume} {81}},\ \bibinfo {pages} {012708} (\bibinfo
  {year} {2010})}\BibitemShut {NoStop}%
\bibitem [{\citenamefont {Harter}\ and\ \citenamefont
  {Denschlag}(2014)}]{harter2014}%
  \BibitemOpen
  \bibfield  {author} {\bibinfo {author} {\bibfnamefont {A.}~\bibnamefont
  {Harter}}\ and\ \bibinfo {author} {\bibfnamefont {J.~H.}\ \bibnamefont
  {Denschlag}},\ }\href@noop {} {\bibfield  {journal} {\bibinfo  {journal}
  {Contemporary Physics}\ }\textbf {\bibinfo {volume} {55}},\ \bibinfo {pages}
  {33} (\bibinfo {year} {2014})}\BibitemShut {NoStop}%
\bibitem [{\citenamefont {Robin~Cote}\ and\ \citenamefont
  {Lukin}(2002)}]{cote2002}%
  \BibitemOpen
  \bibfield  {author} {\bibinfo {author} {\bibfnamefont {V.~K.}\ \bibnamefont
  {Robin~Cote}}\ and\ \bibinfo {author} {\bibfnamefont {M.~D.}\ \bibnamefont
  {Lukin}},\ }\href@noop {} {\bibfield  {journal} {\bibinfo  {journal}
  {Physical review letters}\ }\textbf {\bibinfo {volume} {89}},\ \bibinfo
  {pages} {093001} (\bibinfo {year} {2002})}\BibitemShut {NoStop}%
\bibitem [{\citenamefont {Sayfutyarova}\ \emph {et~al.}(2013)\citenamefont
  {Sayfutyarova}, \citenamefont {Buchachenko}, \citenamefont {Yakovleva},\ and\
  \citenamefont {Belyaev}}]{sayfutyarova2013}%
  \BibitemOpen
  \bibfield  {author} {\bibinfo {author} {\bibfnamefont {E.~R.}\ \bibnamefont
  {Sayfutyarova}}, \bibinfo {author} {\bibfnamefont {A.~A.}\ \bibnamefont
  {Buchachenko}}, \bibinfo {author} {\bibfnamefont {S.~A.}\ \bibnamefont
  {Yakovleva}}, \ and\ \bibinfo {author} {\bibfnamefont {A.~K.}\ \bibnamefont
  {Belyaev}},\ }\href@noop {} {\bibfield  {journal} {\bibinfo  {journal}
  {Physical Review A}\ }\textbf {\bibinfo {volume} {87}},\ \bibinfo {pages}
  {052717} (\bibinfo {year} {2013})}\BibitemShut {NoStop}%
\bibitem [{\citenamefont {Schneider}\ \emph {et~al.}(2010)\citenamefont
  {Schneider}, \citenamefont {Roth}, \citenamefont {Duncker}, \citenamefont
  {Ernsting},\ and\ \citenamefont {Schiller}}]{schneider2010}%
  \BibitemOpen
  \bibfield  {author} {\bibinfo {author} {\bibfnamefont {T.}~\bibnamefont
  {Schneider}}, \bibinfo {author} {\bibfnamefont {B.}~\bibnamefont {Roth}},
  \bibinfo {author} {\bibfnamefont {H.}~\bibnamefont {Duncker}}, \bibinfo
  {author} {\bibfnamefont {I.}~\bibnamefont {Ernsting}}, \ and\ \bibinfo
  {author} {\bibfnamefont {S.}~\bibnamefont {Schiller}},\ }\href@noop {}
  {\bibfield  {journal} {\bibinfo  {journal} {Nature Physics}\ }\textbf
  {\bibinfo {volume} {6}},\ \bibinfo {pages} {275} (\bibinfo {year}
  {2010})}\BibitemShut {NoStop}%
\bibitem [{\citenamefont {Staanum}\ \emph {et~al.}(2010)\citenamefont
  {Staanum}, \citenamefont {Hojbjerre}, \citenamefont {Skyt}, \citenamefont
  {Hansen},\ and\ \citenamefont {Drewsen}}]{staanum2010}%
  \BibitemOpen
  \bibfield  {author} {\bibinfo {author} {\bibfnamefont {P.~F.}\ \bibnamefont
  {Staanum}}, \bibinfo {author} {\bibfnamefont {K.}~\bibnamefont {Hojbjerre}},
  \bibinfo {author} {\bibfnamefont {P.~S.}\ \bibnamefont {Skyt}}, \bibinfo
  {author} {\bibfnamefont {A.~K.}\ \bibnamefont {Hansen}}, \ and\ \bibinfo
  {author} {\bibfnamefont {M.}~\bibnamefont {Drewsen}},\ }\href@noop {}
  {\bibfield  {journal} {\bibinfo  {journal} {Nature Physics}\ }\textbf
  {\bibinfo {volume} {6}},\ \bibinfo {pages} {271} (\bibinfo {year}
  {2010})}\BibitemShut {NoStop}%
\bibitem [{\citenamefont {Langevin}(2009)}]{langevin}%
  \BibitemOpen
  \bibfield  {author} {\bibinfo {author} {\bibfnamefont {P.}~\bibnamefont
  {Langevin}},\ }\href@noop {} {\bibfield  {journal} {\bibinfo  {journal} {Ann.
  Chi. Phys.}\ }\textbf {\bibinfo {volume} {5}},\ \bibinfo {pages} {245}
  (\bibinfo {year} {2009})},\ \bibinfo {note} {a translation is published in
  Appendix II of E. w. McDaniel, Collision Phenomena in Ionized Gases,Wiley,
  New York, 1964}\BibitemShut {NoStop}%
\bibitem [{\citenamefont {Margenau}(1941)}]{marg}%
  \BibitemOpen
  \bibfield  {author} {\bibinfo {author} {\bibfnamefont {H.}~\bibnamefont
  {Margenau}},\ }\href@noop {} {\bibfield  {journal} {\bibinfo  {journal}
  {Philosophy of Science}\ }\textbf {\bibinfo {volume} {8}},\ \bibinfo {pages}
  {603} (\bibinfo {year} {1941})}\BibitemShut {NoStop}%
\bibitem [{\citenamefont {Koutselos}\ and\ \citenamefont
  {Mason}(1986)}]{koutselos1986}%
  \BibitemOpen
  \bibfield  {author} {\bibinfo {author} {\bibfnamefont {A.}~\bibnamefont
  {Koutselos}}\ and\ \bibinfo {author} {\bibfnamefont {E.}~\bibnamefont
  {Mason}},\ }\href@noop {} {\bibfield  {journal} {\bibinfo  {journal} {The
  Journal of chemical physics}\ }\textbf {\bibinfo {volume} {85}},\ \bibinfo
  {pages} {2154} (\bibinfo {year} {1986})}\BibitemShut {NoStop}%
\bibitem [{\citenamefont {Ahlrichs}\ \emph {et~al.}(1988)\citenamefont
  {Ahlrichs}, \citenamefont {Bohm}, \citenamefont {Brode}, \citenamefont
  {Tang},\ and\ \citenamefont {Toennies}}]{ahlrichs1988}%
  \BibitemOpen
  \bibfield  {author} {\bibinfo {author} {\bibfnamefont {R.}~\bibnamefont
  {Ahlrichs}}, \bibinfo {author} {\bibfnamefont {H.}~\bibnamefont {Bohm}},
  \bibinfo {author} {\bibfnamefont {S.}~\bibnamefont {Brode}}, \bibinfo
  {author} {\bibfnamefont {K.}~\bibnamefont {Tang}}, \ and\ \bibinfo {author}
  {\bibfnamefont {J.~P.}\ \bibnamefont {Toennies}},\ }\href@noop {} {\bibfield
  {journal} {\bibinfo  {journal} {The Journal of chemical physics}\ }\textbf
  {\bibinfo {volume} {88}},\ \bibinfo {pages} {6290} (\bibinfo {year}
  {1988})}\BibitemShut {NoStop}%
\bibitem [{\citenamefont {Koutselos}\ \emph {et~al.}(1990)\citenamefont
  {Koutselos}, \citenamefont {Mason},\ and\ \citenamefont
  {Viehland}}]{koutselos1990}%
  \BibitemOpen
  \bibfield  {author} {\bibinfo {author} {\bibfnamefont {A.}~\bibnamefont
  {Koutselos}}, \bibinfo {author} {\bibfnamefont {E.}~\bibnamefont {Mason}}, \
  and\ \bibinfo {author} {\bibfnamefont {L.}~\bibnamefont {Viehland}},\
  }\href@noop {} {\bibfield  {journal} {\bibinfo  {journal} {The Journal of
  chemical physics}\ }\textbf {\bibinfo {volume} {93}},\ \bibinfo {pages}
  {7125} (\bibinfo {year} {1990})}\BibitemShut {NoStop}%
\bibitem [{\citenamefont {Mitroy}\ and\ \citenamefont
  {Zhang}(2008{\natexlab{a}})}]{mitroy2008}%
  \BibitemOpen
  \bibfield  {author} {\bibinfo {author} {\bibfnamefont {J.}~\bibnamefont
  {Mitroy}}\ and\ \bibinfo {author} {\bibfnamefont {J.-Y.}\ \bibnamefont
  {Zhang}},\ }\href@noop {} {\bibfield  {journal} {\bibinfo  {journal} {The
  European Physical Journal D}\ }\textbf {\bibinfo {volume} {46}},\ \bibinfo
  {pages} {415} (\bibinfo {year} {2008}{\natexlab{a}})}\BibitemShut {NoStop}%
\bibitem [{\citenamefont {Tang}\ \emph {et~al.}(2010)\citenamefont {Tang},
  \citenamefont {Zhang}, \citenamefont {Yan}, \citenamefont {Shi},\ and\
  \citenamefont {Mitroy}}]{tang2010}%
  \BibitemOpen
  \bibfield  {author} {\bibinfo {author} {\bibfnamefont {L.-Y.}\ \bibnamefont
  {Tang}}, \bibinfo {author} {\bibfnamefont {J.-Y.}\ \bibnamefont {Zhang}},
  \bibinfo {author} {\bibfnamefont {Z.-C.}\ \bibnamefont {Yan}}, \bibinfo
  {author} {\bibfnamefont {T.-Y.}\ \bibnamefont {Shi}}, \ and\ \bibinfo
  {author} {\bibfnamefont {J.}~\bibnamefont {Mitroy}},\ }\href@noop {}
  {\bibfield  {journal} {\bibinfo  {journal} {The Journal of chemical physics}\
  }\textbf {\bibinfo {volume} {133}},\ \bibinfo {pages} {104306} (\bibinfo
  {year} {2010})}\BibitemShut {NoStop}%
\bibitem [{\citenamefont {Singh}\ \emph {et~al.}(2016)\citenamefont {Singh},
  \citenamefont {Kaur}, \citenamefont {Sahoo},\ and\ \citenamefont
  {Arora}}]{singh2016}%
  \BibitemOpen
  \bibfield  {author} {\bibinfo {author} {\bibfnamefont {S.}~\bibnamefont
  {Singh}}, \bibinfo {author} {\bibfnamefont {K.}~\bibnamefont {Kaur}},
  \bibinfo {author} {\bibfnamefont {B.}~\bibnamefont {Sahoo}}, \ and\ \bibinfo
  {author} {\bibfnamefont {B.}~\bibnamefont {Arora}},\ }\href@noop {}
  {\bibfield  {journal} {\bibinfo  {journal} {arXiv preprint arXiv:1605.05015}\
  } (\bibinfo {year} {2016})}\BibitemShut {NoStop}%
\bibitem [{\citenamefont {Gould}\ and\ \citenamefont
  {Bucko}(2016)}]{gould2016c}%
  \BibitemOpen
  \bibfield  {author} {\bibinfo {author} {\bibfnamefont {T.}~\bibnamefont
  {Gould}}\ and\ \bibinfo {author} {\bibfnamefont {T.}~\bibnamefont {Bucko}},\
  }\href@noop {} {\bibfield  {journal} {\bibinfo  {journal} {Journal of
  chemical theory and computation}\ }\textbf {\bibinfo {volume} {12}},\
  \bibinfo {pages} {3603} (\bibinfo {year} {2016})}\BibitemShut {NoStop}%
\bibitem [{\citenamefont {Grier}\ \emph {et~al.}(2009)\citenamefont {Grier},
  \citenamefont {Cetina}, \citenamefont {Oru{\v{c}}evi{\'c}},\ and\
  \citenamefont {Vuleti{\'c}}}]{grier2009}%
  \BibitemOpen
  \bibfield  {author} {\bibinfo {author} {\bibfnamefont {A.~T.}\ \bibnamefont
  {Grier}}, \bibinfo {author} {\bibfnamefont {M.}~\bibnamefont {Cetina}},
  \bibinfo {author} {\bibfnamefont {F.}~\bibnamefont {Oru{\v{c}}evi{\'c}}}, \
  and\ \bibinfo {author} {\bibfnamefont {V.}~\bibnamefont {Vuleti{\'c}}},\
  }\href@noop {} {\bibfield  {journal} {\bibinfo  {journal} {Physical review
  letters}\ }\textbf {\bibinfo {volume} {102}},\ \bibinfo {pages} {223201}
  (\bibinfo {year} {2009})}\BibitemShut {NoStop}%
\bibitem [{\citenamefont {Zipkes}\ \emph
  {et~al.}(2010{\natexlab{b}})\citenamefont {Zipkes}, \citenamefont {Palzer},
  \citenamefont {Ratschbacher}, \citenamefont {Sias},\ and\ \citenamefont
  {Kohl}}]{zipkes2010cold}%
  \BibitemOpen
  \bibfield  {author} {\bibinfo {author} {\bibfnamefont {C.}~\bibnamefont
  {Zipkes}}, \bibinfo {author} {\bibfnamefont {S.}~\bibnamefont {Palzer}},
  \bibinfo {author} {\bibfnamefont {L.}~\bibnamefont {Ratschbacher}}, \bibinfo
  {author} {\bibfnamefont {C.}~\bibnamefont {Sias}}, \ and\ \bibinfo {author}
  {\bibfnamefont {M.}~\bibnamefont {Kohl}},\ }\href@noop {} {\bibfield
  {journal} {\bibinfo  {journal} {Physical review letters}\ }\textbf {\bibinfo
  {volume} {105}},\ \bibinfo {pages} {133201} (\bibinfo {year}
  {2010}{\natexlab{b}})}\BibitemShut {NoStop}%
\bibitem [{\citenamefont {Zipkes}\ \emph
  {et~al.}(2010{\natexlab{c}})\citenamefont {Zipkes}, \citenamefont {Palzer},
  \citenamefont {Sias},\ and\ \citenamefont {K{\"o}hl}}]{zipkes2010trapped}%
  \BibitemOpen
  \bibfield  {author} {\bibinfo {author} {\bibfnamefont {C.}~\bibnamefont
  {Zipkes}}, \bibinfo {author} {\bibfnamefont {S.}~\bibnamefont {Palzer}},
  \bibinfo {author} {\bibfnamefont {C.}~\bibnamefont {Sias}}, \ and\ \bibinfo
  {author} {\bibfnamefont {M.}~\bibnamefont {K{\"o}hl}},\ }\href@noop {}
  {\bibfield  {journal} {\bibinfo  {journal} {Nature}\ }\textbf {\bibinfo
  {volume} {464}},\ \bibinfo {pages} {388} (\bibinfo {year}
  {2010}{\natexlab{c}})}\BibitemShut {NoStop}%
\bibitem [{\citenamefont {Hall}\ \emph {et~al.}(2011)\citenamefont {Hall},
  \citenamefont {Aymar}, \citenamefont {Bouloufa-Maafa}, \citenamefont
  {Dulieu},\ and\ \citenamefont {Willitsch}}]{hall2011light}%
  \BibitemOpen
  \bibfield  {author} {\bibinfo {author} {\bibfnamefont {F.~H.}\ \bibnamefont
  {Hall}}, \bibinfo {author} {\bibfnamefont {M.}~\bibnamefont {Aymar}},
  \bibinfo {author} {\bibfnamefont {N.}~\bibnamefont {Bouloufa-Maafa}},
  \bibinfo {author} {\bibfnamefont {O.}~\bibnamefont {Dulieu}}, \ and\ \bibinfo
  {author} {\bibfnamefont {S.}~\bibnamefont {Willitsch}},\ }\href@noop {}
  {\bibfield  {journal} {\bibinfo  {journal} {Physical review letters}\
  }\textbf {\bibinfo {volume} {107}},\ \bibinfo {pages} {243202} (\bibinfo
  {year} {2011})}\BibitemShut {NoStop}%
\bibitem [{\citenamefont {Hall}\ and\ \citenamefont
  {Willitsch}(2012)}]{hall2012millikelvin}%
  \BibitemOpen
  \bibfield  {author} {\bibinfo {author} {\bibfnamefont {F.~H.}\ \bibnamefont
  {Hall}}\ and\ \bibinfo {author} {\bibfnamefont {S.}~\bibnamefont
  {Willitsch}},\ }\href@noop {} {\bibfield  {journal} {\bibinfo  {journal}
  {Physical review letters}\ }\textbf {\bibinfo {volume} {109}},\ \bibinfo
  {pages} {233202} (\bibinfo {year} {2012})}\BibitemShut {NoStop}%
\bibitem [{\citenamefont {Sullivan}\ \emph {et~al.}(2012)\citenamefont
  {Sullivan}, \citenamefont {Rellergert}, \citenamefont {Kotochigova},\ and\
  \citenamefont {Hudson}}]{sullivan2012}%
  \BibitemOpen
  \bibfield  {author} {\bibinfo {author} {\bibfnamefont {S.~T.}\ \bibnamefont
  {Sullivan}}, \bibinfo {author} {\bibfnamefont {W.~G.}\ \bibnamefont
  {Rellergert}}, \bibinfo {author} {\bibfnamefont {S.}~\bibnamefont
  {Kotochigova}}, \ and\ \bibinfo {author} {\bibfnamefont {E.~R.}\ \bibnamefont
  {Hudson}},\ }\href@noop {} {\bibfield  {journal} {\bibinfo  {journal}
  {Physical review letters}\ }\textbf {\bibinfo {volume} {109}},\ \bibinfo
  {pages} {223002} (\bibinfo {year} {2012})}\BibitemShut {NoStop}%
\bibitem [{\citenamefont {Ravi}\ \emph {et~al.}(2012)\citenamefont {Ravi},
  \citenamefont {Lee}, \citenamefont {Sharma}, \citenamefont {Werth},\ and\
  \citenamefont {Rangwala}}]{ravi2012}%
  \BibitemOpen
  \bibfield  {author} {\bibinfo {author} {\bibfnamefont {K.}~\bibnamefont
  {Ravi}}, \bibinfo {author} {\bibfnamefont {S.}~\bibnamefont {Lee}}, \bibinfo
  {author} {\bibfnamefont {A.}~\bibnamefont {Sharma}}, \bibinfo {author}
  {\bibfnamefont {G.}~\bibnamefont {Werth}}, \ and\ \bibinfo {author}
  {\bibfnamefont {S.}~\bibnamefont {Rangwala}},\ }\href@noop {} {\bibfield
  {journal} {\bibinfo  {journal} {Nature communications}\ }\textbf {\bibinfo
  {volume} {3}},\ \bibinfo {pages} {1} (\bibinfo {year} {2012})}\BibitemShut
  {NoStop}%
\bibitem [{\citenamefont {Lee}\ \emph {et~al.}(2013)\citenamefont {Lee},
  \citenamefont {Ravi},\ and\ \citenamefont {Rangwala}}]{lee2013}%
  \BibitemOpen
  \bibfield  {author} {\bibinfo {author} {\bibfnamefont {S.}~\bibnamefont
  {Lee}}, \bibinfo {author} {\bibfnamefont {K.}~\bibnamefont {Ravi}}, \ and\
  \bibinfo {author} {\bibfnamefont {S.}~\bibnamefont {Rangwala}},\ }\href@noop
  {} {\bibfield  {journal} {\bibinfo  {journal} {Physical Review A}\ }\textbf
  {\bibinfo {volume} {87}},\ \bibinfo {pages} {052701} (\bibinfo {year}
  {2013})}\BibitemShut {NoStop}%
\bibitem [{\citenamefont {Tacconi}\ \emph {et~al.}(2011)\citenamefont
  {Tacconi}, \citenamefont {Gianturco},\ and\ \citenamefont
  {Belyaev}}]{tacconi2011}%
  \BibitemOpen
  \bibfield  {author} {\bibinfo {author} {\bibfnamefont {M.}~\bibnamefont
  {Tacconi}}, \bibinfo {author} {\bibfnamefont {F.}~\bibnamefont {Gianturco}},
  \ and\ \bibinfo {author} {\bibfnamefont {A.}~\bibnamefont {Belyaev}},\
  }\href@noop {} {\bibfield  {journal} {\bibinfo  {journal} {Physical Chemistry
  Chemical Physics}\ }\textbf {\bibinfo {volume} {13}},\ \bibinfo {pages}
  {19156} (\bibinfo {year} {2011})}\BibitemShut {NoStop}%
\bibitem [{\citenamefont {McLaughlin}\ \emph {et~al.}(2014)\citenamefont
  {McLaughlin}, \citenamefont {Lamb}, \citenamefont {Lane},\ and\ \citenamefont
  {McCann}}]{mclaughlin2014}%
  \BibitemOpen
  \bibfield  {author} {\bibinfo {author} {\bibfnamefont {B.}~\bibnamefont
  {McLaughlin}}, \bibinfo {author} {\bibfnamefont {H.}~\bibnamefont {Lamb}},
  \bibinfo {author} {\bibfnamefont {I.}~\bibnamefont {Lane}}, \ and\ \bibinfo
  {author} {\bibfnamefont {J.}~\bibnamefont {McCann}},\ }\href@noop {}
  {\bibfield  {journal} {\bibinfo  {journal} {Journal of Physics B: Atomic,
  Molecular and Optical Physics}\ }\textbf {\bibinfo {volume} {47}},\ \bibinfo
  {pages} {145201} (\bibinfo {year} {2014})}\BibitemShut {NoStop}%
\bibitem [{\citenamefont {Axilrod}\ and\ \citenamefont
  {Teller}(1943)}]{axilrod1943}%
  \BibitemOpen
  \bibfield  {author} {\bibinfo {author} {\bibfnamefont {B.}~\bibnamefont
  {Axilrod}}\ and\ \bibinfo {author} {\bibfnamefont {E.}~\bibnamefont
  {Teller}},\ }\href@noop {} {\bibfield  {journal} {\bibinfo  {journal} {The
  Journal of Chemical Physics}\ }\textbf {\bibinfo {volume} {11}},\ \bibinfo
  {pages} {299} (\bibinfo {year} {1943})}\BibitemShut {NoStop}%
\bibitem [{\citenamefont {Muto}(1943)}]{muto1943}%
  \BibitemOpen
  \bibfield  {author} {\bibinfo {author} {\bibfnamefont {Y.}~\bibnamefont
  {Muto}},\ }\href@noop {} {\bibfield  {journal} {\bibinfo  {journal} {J. Phys.
  Math. Soc. Jpn}\ }\textbf {\bibinfo {volume} {17}},\ \bibinfo {pages} {629}
  (\bibinfo {year} {1943})}\BibitemShut {NoStop}%
\bibitem [{\citenamefont {Axilrod}(1951)}]{axilrod1951}%
  \BibitemOpen
  \bibfield  {author} {\bibinfo {author} {\bibfnamefont {B.~M.}\ \bibnamefont
  {Axilrod}},\ }\href@noop {} {\bibfield  {journal} {\bibinfo  {journal} {The
  Journal of Chemical Physics}\ }\textbf {\bibinfo {volume} {19}},\ \bibinfo
  {pages} {719} (\bibinfo {year} {1951})}\BibitemShut {NoStop}%
\bibitem [{\citenamefont {Marinescu}\ and\ \citenamefont
  {Starace}(1997{\natexlab{a}})}]{marinescu1997}%
  \BibitemOpen
  \bibfield  {author} {\bibinfo {author} {\bibfnamefont {M.}~\bibnamefont
  {Marinescu}}\ and\ \bibinfo {author} {\bibfnamefont {A.~F.}\ \bibnamefont
  {Starace}},\ }\href@noop {} {\bibfield  {journal} {\bibinfo  {journal}
  {Physical Review A}\ }\textbf {\bibinfo {volume} {55}},\ \bibinfo {pages}
  {2067} (\bibinfo {year} {1997}{\natexlab{a}})}\BibitemShut {NoStop}%
\bibitem [{\citenamefont {Patil}\ and\ \citenamefont {Tang}(1997)}]{patil1997}%
  \BibitemOpen
  \bibfield  {author} {\bibinfo {author} {\bibfnamefont {S.~H.}\ \bibnamefont
  {Patil}}\ and\ \bibinfo {author} {\bibfnamefont {K.~T.}\ \bibnamefont
  {Tang}},\ }\href@noop {} {\bibfield  {journal} {\bibinfo  {journal} {The
  Journal of Chemical Physics}\ }\textbf {\bibinfo {volume} {106}},\ \bibinfo
  {pages} {2298} (\bibinfo {year} {1997})}\BibitemShut {NoStop}%
\bibitem [{\citenamefont {Anatole~von Lilienfeld}\ and\ \citenamefont
  {Tkatchenko}(2010)}]{anatole2010}%
  \BibitemOpen
  \bibfield  {author} {\bibinfo {author} {\bibfnamefont {O.}~\bibnamefont
  {Anatole~von Lilienfeld}}\ and\ \bibinfo {author} {\bibfnamefont
  {A.}~\bibnamefont {Tkatchenko}},\ }\href@noop {} {\bibfield  {journal}
  {\bibinfo  {journal} {The Journal of chemical physics}\ }\textbf {\bibinfo
  {volume} {132}},\ \bibinfo {pages} {234109} (\bibinfo {year}
  {2010})}\BibitemShut {NoStop}%
\bibitem [{\citenamefont {Huang}\ and\ \citenamefont {Sun}(2011)}]{huang2011}%
  \BibitemOpen
  \bibfield  {author} {\bibinfo {author} {\bibfnamefont {S.-Z.}\ \bibnamefont
  {Huang}}\ and\ \bibinfo {author} {\bibfnamefont {Q.-F.}\ \bibnamefont
  {Sun}},\ }\href@noop {} {\bibfield  {journal} {\bibinfo  {journal} {The
  Journal of chemical physics}\ }\textbf {\bibinfo {volume} {134}},\ \bibinfo
  {pages} {144110} (\bibinfo {year} {2011})}\BibitemShut {NoStop}%
\bibitem [{\citenamefont {Gao}\ and\ \citenamefont {Starace}(1989)}]{gao1989}%
  \BibitemOpen
  \bibfield  {author} {\bibinfo {author} {\bibfnamefont {B.}~\bibnamefont
  {Gao}}\ and\ \bibinfo {author} {\bibfnamefont {A.~F.}\ \bibnamefont
  {Starace}},\ }\href@noop {} {\bibfield  {journal} {\bibinfo  {journal}
  {Physical Review A}\ }\textbf {\bibinfo {volume} {39}},\ \bibinfo {pages}
  {4550} (\bibinfo {year} {1989})}\BibitemShut {NoStop}%
\bibitem [{\citenamefont {Zhang}\ \emph {et~al.}(2020)\citenamefont {Zhang},
  \citenamefont {Cheng}, \citenamefont {Zhang},\ and\ \citenamefont
  {Zhai}}]{zhang2020}%
  \BibitemOpen
  \bibfield  {author} {\bibinfo {author} {\bibfnamefont {R.}~\bibnamefont
  {Zhang}}, \bibinfo {author} {\bibfnamefont {Y.}~\bibnamefont {Cheng}},
  \bibinfo {author} {\bibfnamefont {P.}~\bibnamefont {Zhang}}, \ and\ \bibinfo
  {author} {\bibfnamefont {H.}~\bibnamefont {Zhai}},\ }\href@noop {} {\bibfield
   {journal} {\bibinfo  {journal} {Nature Reviews Physics}\ ,\ \bibinfo {pages}
  {1}} (\bibinfo {year} {2020})}\BibitemShut {NoStop}%
\bibitem [{\citenamefont {Zhang}\ \emph {et~al.}(2019)\citenamefont {Zhang},
  \citenamefont {Cheng}, \citenamefont {Zhang},\ and\ \citenamefont
  {Zhai}}]{zhang2019}%
  \BibitemOpen
  \bibfield  {author} {\bibinfo {author} {\bibfnamefont {R.}~\bibnamefont
  {Zhang}}, \bibinfo {author} {\bibfnamefont {Y.}~\bibnamefont {Cheng}},
  \bibinfo {author} {\bibfnamefont {P.}~\bibnamefont {Zhang}}, \ and\ \bibinfo
  {author} {\bibfnamefont {H.}~\bibnamefont {Zhai}},\ }\href@noop {} {\bibfield
   {journal} {\bibinfo  {journal} {arXiv preprint arXiv:1908.04973}\ }
  (\bibinfo {year} {2019})}\BibitemShut {NoStop}%
\bibitem [{\citenamefont {Wineland}(2013)}]{wineland2013}%
  \BibitemOpen
  \bibfield  {author} {\bibinfo {author} {\bibfnamefont {D.~J.}\ \bibnamefont
  {Wineland}},\ }\href@noop {} {\bibfield  {journal} {\bibinfo  {journal}
  {Reviews of Modern Physics}\ }\textbf {\bibinfo {volume} {85}},\ \bibinfo
  {pages} {1103} (\bibinfo {year} {2013})}\BibitemShut {NoStop}%
\bibitem [{\citenamefont {Haroche}(2013)}]{haroche2013}%
  \BibitemOpen
  \bibfield  {author} {\bibinfo {author} {\bibfnamefont {S.}~\bibnamefont
  {Haroche}},\ }\href@noop {} {\bibfield  {journal} {\bibinfo  {journal}
  {Reviews of Modern Physics}\ }\textbf {\bibinfo {volume} {85}},\ \bibinfo
  {pages} {1083} (\bibinfo {year} {2013})}\BibitemShut {NoStop}%
\bibitem [{\citenamefont {C{\^o}t{\'e}}(2016)}]{cote2016}%
  \BibitemOpen
  \bibfield  {author} {\bibinfo {author} {\bibfnamefont {R.}~\bibnamefont
  {C{\^o}t{\'e}}},\ }in\ \href@noop {} {\emph {\bibinfo {booktitle} {Advances
  In Atomic, Molecular, and Optical Physics}}},\ Vol.~\bibinfo {volume} {65}\
  (\bibinfo  {publisher} {Elsevier},\ \bibinfo {year} {2016})\ pp.\ \bibinfo
  {pages} {67--126}\BibitemShut {NoStop}%
\bibitem [{\citenamefont {Tomza}\ \emph {et~al.}(2019)\citenamefont {Tomza},
  \citenamefont {Jachymski}, \citenamefont {Gerritsma}, \citenamefont
  {Negretti}, \citenamefont {Calarco}, \citenamefont {Idziaszek},\ and\
  \citenamefont {Julienne}}]{tomza2019}%
  \BibitemOpen
  \bibfield  {author} {\bibinfo {author} {\bibfnamefont {M.}~\bibnamefont
  {Tomza}}, \bibinfo {author} {\bibfnamefont {K.}~\bibnamefont {Jachymski}},
  \bibinfo {author} {\bibfnamefont {R.}~\bibnamefont {Gerritsma}}, \bibinfo
  {author} {\bibfnamefont {A.}~\bibnamefont {Negretti}}, \bibinfo {author}
  {\bibfnamefont {T.}~\bibnamefont {Calarco}}, \bibinfo {author} {\bibfnamefont
  {Z.}~\bibnamefont {Idziaszek}}, \ and\ \bibinfo {author} {\bibfnamefont
  {P.~S.}\ \bibnamefont {Julienne}},\ }\href@noop {} {\bibfield  {journal}
  {\bibinfo  {journal} {Reviews of Modern Physics}\ }\textbf {\bibinfo {volume}
  {91}},\ \bibinfo {pages} {035001} (\bibinfo {year} {2019})}\BibitemShut
  {NoStop}%
\bibitem [{\citenamefont {{\'S}mia{\l}kowski}\ and\ \citenamefont
  {Tomza}(2020)}]{smialkowski2020}%
  \BibitemOpen
  \bibfield  {author} {\bibinfo {author} {\bibfnamefont {M.}~\bibnamefont
  {{\'S}mia{\l}kowski}}\ and\ \bibinfo {author} {\bibfnamefont
  {M.}~\bibnamefont {Tomza}},\ }\href@noop {} {\bibfield  {journal} {\bibinfo
  {journal} {Physical Review A}\ }\textbf {\bibinfo {volume} {101}},\ \bibinfo
  {pages} {012501} (\bibinfo {year} {2020})}\BibitemShut {NoStop}%
\bibitem [{\citenamefont {Li}\ and\ \citenamefont {Stwalley}(1973)}]{li1973}%
  \BibitemOpen
  \bibfield  {author} {\bibinfo {author} {\bibfnamefont {K.~C.}\ \bibnamefont
  {Li}}\ and\ \bibinfo {author} {\bibfnamefont {W.~C.}\ \bibnamefont
  {Stwalley}},\ }\href@noop {} {\bibfield  {journal} {\bibinfo  {journal} {The
  Journal of Chemical Physics}\ }\textbf {\bibinfo {volume} {59}},\ \bibinfo
  {pages} {4423} (\bibinfo {year} {1973})}\BibitemShut {NoStop}%
\bibitem [{\citenamefont {Allard}\ \emph {et~al.}(2003)\citenamefont {Allard},
  \citenamefont {Samuelis}, \citenamefont {Pashov}, \citenamefont
  {Kn{\"o}ckel},\ and\ \citenamefont {Tiemann}}]{allard2003}%
  \BibitemOpen
  \bibfield  {author} {\bibinfo {author} {\bibfnamefont {O.}~\bibnamefont
  {Allard}}, \bibinfo {author} {\bibfnamefont {C.}~\bibnamefont {Samuelis}},
  \bibinfo {author} {\bibfnamefont {A.}~\bibnamefont {Pashov}}, \bibinfo
  {author} {\bibfnamefont {H.}~\bibnamefont {Kn{\"o}ckel}}, \ and\ \bibinfo
  {author} {\bibfnamefont {E.}~\bibnamefont {Tiemann}},\ }\href@noop {}
  {\bibfield  {journal} {\bibinfo  {journal} {The European Physical Journal
  D-Atomic, Molecular, Optical and Plasma Physics}\ }\textbf {\bibinfo {volume}
  {26}},\ \bibinfo {pages} {155} (\bibinfo {year} {2003})}\BibitemShut
  {NoStop}%
\bibitem [{\citenamefont {Kramida}\ \emph {et~al.}(2018)\citenamefont
  {Kramida}, \citenamefont {{Yu.~Ralchenko}}, \citenamefont {Reader},\ and\
  \citenamefont {{and NIST ASD Team}}}]{NIST}%
  \BibitemOpen
  \bibfield  {author} {\bibinfo {author} {\bibfnamefont {A.}~\bibnamefont
  {Kramida}}, \bibinfo {author} {\bibnamefont {{Yu.~Ralchenko}}}, \bibinfo
  {author} {\bibfnamefont {J.}~\bibnamefont {Reader}}, \ and\ \bibinfo {author}
  {\bibnamefont {{and NIST ASD Team}}},\ }\href@noop {} {}\bibinfo
  {howpublished} {{NIST Atomic Spectra Database (ver. 5.6.1), [Online].
  Available: {\tt{https://physics.nist.gov/asd}} [2019, March 29]. National
  Institute of Standards and Technology, Gaithersburg, MD.}} (\bibinfo {year}
  {2018})\BibitemShut {NoStop}%
\bibitem [{\citenamefont {Risberg}(1955)}]{risberg1955}%
  \BibitemOpen
  \bibfield  {author} {\bibinfo {author} {\bibfnamefont {P.}~\bibnamefont
  {Risberg}},\ }\href@noop {} {\bibfield  {journal} {\bibinfo  {journal} {Arkiv
  for Fysik}\ }\textbf {\bibinfo {volume} {9}},\ \bibinfo {pages} {483}
  (\bibinfo {year} {1955})}\BibitemShut {NoStop}%
\bibitem [{\citenamefont {Edl{\'e}n}\ and\ \citenamefont
  {Risberg}(1956)}]{edlen1956}%
  \BibitemOpen
  \bibfield  {author} {\bibinfo {author} {\bibfnamefont {B.}~\bibnamefont
  {Edl{\'e}n}}\ and\ \bibinfo {author} {\bibfnamefont {P.}~\bibnamefont
  {Risberg}},\ }\href@noop {} {\bibfield  {journal} {\bibinfo  {journal} {Arkiv
  For Fysik}\ }\textbf {\bibinfo {volume} {10}},\ \bibinfo {pages} {553}
  (\bibinfo {year} {1956})}\BibitemShut {NoStop}%
\bibitem [{\citenamefont {Sullivan}(1938)}]{sullivan1938}%
  \BibitemOpen
  \bibfield  {author} {\bibinfo {author} {\bibfnamefont {F.}~\bibnamefont
  {Sullivan}},\ }\href@noop {} {\bibfield  {journal} {\bibinfo  {journal}
  {PhDT}\ } (\bibinfo {year} {1938})}\BibitemShut {NoStop}%
\bibitem [{\citenamefont {Karlsson}\ and\ \citenamefont
  {Litz{\'e}n}(1999)}]{karlsson1999}%
  \BibitemOpen
  \bibfield  {author} {\bibinfo {author} {\bibfnamefont {H.}~\bibnamefont
  {Karlsson}}\ and\ \bibinfo {author} {\bibfnamefont {U.}~\bibnamefont
  {Litz{\'e}n}},\ }\href@noop {} {\bibfield  {journal} {\bibinfo  {journal}
  {Physica Scripta}\ }\textbf {\bibinfo {volume} {60}},\ \bibinfo {pages} {321}
  (\bibinfo {year} {1999})}\BibitemShut {NoStop}%
\bibitem [{\citenamefont {Meissner}(1938)}]{meissner1938}%
  \BibitemOpen
  \bibfield  {author} {\bibinfo {author} {\bibfnamefont {K.}~\bibnamefont
  {Meissner}},\ }\href@noop {} {\bibfield  {journal} {\bibinfo  {journal}
  {Annalen der Physik}\ }\textbf {\bibinfo {volume} {423}},\ \bibinfo {pages}
  {505} (\bibinfo {year} {1938})}\BibitemShut {NoStop}%
\bibitem [{\citenamefont {Risberg}(1965)}]{risberg1965}%
  \BibitemOpen
  \bibfield  {author} {\bibinfo {author} {\bibfnamefont {G.}~\bibnamefont
  {Risberg}},\ }\href@noop {} {\bibfield  {journal} {\bibinfo  {journal} {Arkiv
  for Fysik}\ }\textbf {\bibinfo {volume} {28}},\ \bibinfo {pages} {381}
  (\bibinfo {year} {1965})}\BibitemShut {NoStop}%
\bibitem [{\citenamefont {Risberg}(1968)}]{risberg1968}%
  \BibitemOpen
  \bibfield  {author} {\bibinfo {author} {\bibfnamefont {G.}~\bibnamefont
  {Risberg}},\ }\href@noop {} {\bibfield  {journal} {\bibinfo  {journal} {Arkiv
  for Fysik}\ }\textbf {\bibinfo {volume} {37}},\ \bibinfo {pages} {231}
  (\bibinfo {year} {1968})}\BibitemShut {NoStop}%
\bibitem [{\citenamefont {Lim}\ and\ \citenamefont
  {Schwerdtfeger}(2004)}]{lim2004}%
  \BibitemOpen
  \bibfield  {author} {\bibinfo {author} {\bibfnamefont {I.~S.}\ \bibnamefont
  {Lim}}\ and\ \bibinfo {author} {\bibfnamefont {P.}~\bibnamefont
  {Schwerdtfeger}},\ }\href@noop {} {\bibfield  {journal} {\bibinfo  {journal}
  {Physical Review A}\ }\textbf {\bibinfo {volume} {70}},\ \bibinfo {pages}
  {062501} (\bibinfo {year} {2004})}\BibitemShut {NoStop}%
\bibitem [{\citenamefont {Porsev}\ and\ \citenamefont
  {Derevianko}(2006{\natexlab{a}})}]{dere-seg}%
  \BibitemOpen
  \bibfield  {author} {\bibinfo {author} {\bibfnamefont {S.~G.}\ \bibnamefont
  {Porsev}}\ and\ \bibinfo {author} {\bibfnamefont {A.}~\bibnamefont
  {Derevianko}},\ }\href@noop {} {\bibfield  {journal} {\bibinfo  {journal}
  {Journal of Experimental and Theoretical Physics}\ }\textbf {\bibinfo
  {volume} {102}},\ \bibinfo {pages} {195} (\bibinfo {year}
  {2006}{\natexlab{a}})}\BibitemShut {NoStop}%
\bibitem [{\citenamefont {Chang}(1983)}]{chang1983}%
  \BibitemOpen
  \bibfield  {author} {\bibinfo {author} {\bibfnamefont {E.}~\bibnamefont
  {Chang}},\ }\href@noop {} {\bibfield  {journal} {\bibinfo  {journal} {Journal
  of Physics B: Atomic and Molecular Physics}\ }\textbf {\bibinfo {volume}
  {16}},\ \bibinfo {pages} {L539} (\bibinfo {year} {1983})}\BibitemShut
  {NoStop}%
\bibitem [{\citenamefont {Nunkaew}\ \emph {et~al.}(2009)\citenamefont
  {Nunkaew}, \citenamefont {Shuman},\ and\ \citenamefont
  {Gallagher}}]{nunkaew2009}%
  \BibitemOpen
  \bibfield  {author} {\bibinfo {author} {\bibfnamefont {J.}~\bibnamefont
  {Nunkaew}}, \bibinfo {author} {\bibfnamefont {E.}~\bibnamefont {Shuman}}, \
  and\ \bibinfo {author} {\bibfnamefont {T.}~\bibnamefont {Gallagher}},\
  }\href@noop {} {\bibfield  {journal} {\bibinfo  {journal} {Physical Review
  A}\ }\textbf {\bibinfo {volume} {79}},\ \bibinfo {pages} {054501} (\bibinfo
  {year} {2009})}\BibitemShut {NoStop}%
\bibitem [{\citenamefont {Snow}\ and\ \citenamefont
  {Lundeen}(2007)}]{snow2007}%
  \BibitemOpen
  \bibfield  {author} {\bibinfo {author} {\bibfnamefont {E.}~\bibnamefont
  {Snow}}\ and\ \bibinfo {author} {\bibfnamefont {S.}~\bibnamefont {Lundeen}},\
  }\href@noop {} {\bibfield  {journal} {\bibinfo  {journal} {Physical Review
  A}\ }\textbf {\bibinfo {volume} {76}},\ \bibinfo {pages} {052505} (\bibinfo
  {year} {2007})}\BibitemShut {NoStop}%
\bibitem [{\citenamefont {L.~Lundin}\ and\ \citenamefont
  {Martinson}(1973)}]{36}%
  \BibitemOpen
  \bibfield  {author} {\bibinfo {author} {\bibfnamefont {J.~H.}\ \bibnamefont
  {L.~Lundin}, \bibfnamefont {B.~Engman}}\ and\ \bibinfo {author}
  {\bibfnamefont {I.}~\bibnamefont {Martinson}},\ }\href@noop {} {\bibfield
  {journal} {\bibinfo  {journal} {Phys. Scr. T}\ }\textbf {\bibinfo {volume}
  {8}},\ \bibinfo {pages} {274} (\bibinfo {year} {1973})}\BibitemShut {NoStop}%
\bibitem [{\citenamefont {Degenhardt}\ \emph {et~al.}(2003)\citenamefont
  {Degenhardt}, \citenamefont {Binnewies},\ and\ \citenamefont {Wilpers}}]{28}%
  \BibitemOpen
  \bibfield  {author} {\bibinfo {author} {\bibfnamefont {C.}~\bibnamefont
  {Degenhardt}}, \bibinfo {author} {\bibfnamefont {T.}~\bibnamefont
  {Binnewies}}, \ and\ \bibinfo {author} {\bibfnamefont {G.}~\bibnamefont
  {Wilpers}},\ }\href@noop {} {\bibfield  {journal} {\bibinfo  {journal} {Phys.
  Rev. A}\ }\textbf {\bibinfo {volume} {67}} (\bibinfo {year}
  {2003})}\BibitemShut {NoStop}%
\bibitem [{\citenamefont {Lurio}\ and\ \citenamefont {Novick}(1964)}]{38}%
  \BibitemOpen
  \bibfield  {author} {\bibinfo {author} {\bibfnamefont {A.}~\bibnamefont
  {Lurio}}\ and\ \bibinfo {author} {\bibfnamefont {R.}~\bibnamefont {Novick}},\
  }\href@noop {} {\bibfield  {journal} {\bibinfo  {journal} {Phys. Scr.}\
  }\textbf {\bibinfo {volume} {134}} (\bibinfo {year} {1964})}\BibitemShut
  {NoStop}%
\bibitem [{\citenamefont {Schwartz}\ \emph {et~al.}(1974)\citenamefont
  {Schwartz}, \citenamefont {Miller},\ and\ \citenamefont
  {Bederson}}]{schwartz1974}%
  \BibitemOpen
  \bibfield  {author} {\bibinfo {author} {\bibfnamefont {H.~L.}\ \bibnamefont
  {Schwartz}}, \bibinfo {author} {\bibfnamefont {T.~M.}\ \bibnamefont
  {Miller}}, \ and\ \bibinfo {author} {\bibfnamefont {B.}~\bibnamefont
  {Bederson}},\ }\href@noop {} {\bibfield  {journal} {\bibinfo  {journal}
  {Physical Review A}\ }\textbf {\bibinfo {volume} {10}},\ \bibinfo {pages}
  {1924} (\bibinfo {year} {1974})}\BibitemShut {NoStop}%
\bibitem [{\citenamefont {Marinescu}\ and\ \citenamefont
  {Starace}(1997{\natexlab{b}})}]{marin}%
  \BibitemOpen
  \bibfield  {author} {\bibinfo {author} {\bibfnamefont {M.}~\bibnamefont
  {Marinescu}}\ and\ \bibinfo {author} {\bibfnamefont {A.~F.}\ \bibnamefont
  {Starace}},\ }\href {\doibase 10.1103/PhysRevA.55.2067} {\bibfield  {journal}
  {\bibinfo  {journal} {Phys. Rev. A}\ }\textbf {\bibinfo {volume} {55}},\
  \bibinfo {pages} {2067} (\bibinfo {year} {1997}{\natexlab{b}})}\BibitemShut
  {NoStop}%
\bibitem [{\citenamefont {Slater}\ and\ \citenamefont {Kirkwood}(1931)}]{sk}%
  \BibitemOpen
  \bibfield  {author} {\bibinfo {author} {\bibfnamefont {J.~C.}\ \bibnamefont
  {Slater}}\ and\ \bibinfo {author} {\bibfnamefont {J.~G.}\ \bibnamefont
  {Kirkwood}},\ }\href@noop {} {\bibfield  {journal} {\bibinfo  {journal}
  {Phys. Rev}\ }\textbf {\bibinfo {volume} {37}},\ \bibinfo {pages} {682}
  (\bibinfo {year} {1931})}\BibitemShut {NoStop}%
\bibitem [{\citenamefont {Kramer}\ and\ \citenamefont
  {Herschbach}(1970)}]{kramer}%
  \BibitemOpen
  \bibfield  {author} {\bibinfo {author} {\bibfnamefont {H.}~\bibnamefont
  {Kramer}}\ and\ \bibinfo {author} {\bibfnamefont {D.}~\bibnamefont
  {Herschbach}},\ }\href@noop {} {\bibfield  {journal} {\bibinfo  {journal}
  {The Journal of Chemical Physics}\ }\textbf {\bibinfo {volume} {53}},\
  \bibinfo {pages} {2792} (\bibinfo {year} {1970})}\BibitemShut {NoStop}%
\bibitem [{\citenamefont {Midzuno}\ and\ \citenamefont
  {Kihara}(1956)}]{midzuno}%
  \BibitemOpen
  \bibfield  {author} {\bibinfo {author} {\bibfnamefont {Y.}~\bibnamefont
  {Midzuno}}\ and\ \bibinfo {author} {\bibfnamefont {T.}~\bibnamefont
  {Kihara}},\ }\href@noop {} {\bibfield  {journal} {\bibinfo  {journal}
  {Journal of the Physical Society of Japan}\ }\textbf {\bibinfo {volume}
  {11}},\ \bibinfo {pages} {1045} (\bibinfo {year} {1956})}\BibitemShut
  {NoStop}%
\bibitem [{\citenamefont {Tang}(1969)}]{tang1969}%
  \BibitemOpen
  \bibfield  {author} {\bibinfo {author} {\bibfnamefont {K.}~\bibnamefont
  {Tang}},\ }\href@noop {} {\bibfield  {journal} {\bibinfo  {journal} {Physical
  Review}\ }\textbf {\bibinfo {volume} {177}},\ \bibinfo {pages} {108}
  (\bibinfo {year} {1969})}\BibitemShut {NoStop}%
\bibitem [{\citenamefont {Barklem}\ and\ \citenamefont {Omara}(2000)}]{49}%
  \BibitemOpen
  \bibfield  {author} {\bibinfo {author} {\bibfnamefont {P.~S.}\ \bibnamefont
  {Barklem}}\ and\ \bibinfo {author} {\bibfnamefont {B.~J.}\ \bibnamefont
  {Omara}},\ }\href@noop {} {\bibfield  {journal} {\bibinfo  {journal} {Mon.
  Not. R. Astron. Soc.}\ }\textbf {\bibinfo {volume} {311}},\ \bibinfo {pages}
  {535} (\bibinfo {year} {2000})}\BibitemShut {NoStop}%
\bibitem [{\citenamefont {Arora}\ \emph {et~al.}(2012)\citenamefont {Arora},
  \citenamefont {Nandy},\ and\ \citenamefont {Sahoo}}]{arora2012}%
  \BibitemOpen
  \bibfield  {author} {\bibinfo {author} {\bibfnamefont {B.}~\bibnamefont
  {Arora}}, \bibinfo {author} {\bibfnamefont {D.}~\bibnamefont {Nandy}}, \ and\
  \bibinfo {author} {\bibfnamefont {B.}~\bibnamefont {Sahoo}},\ }\href@noop {}
  {\bibfield  {journal} {\bibinfo  {journal} {Physical Review A}\ }\textbf
  {\bibinfo {volume} {85}},\ \bibinfo {pages} {012506} (\bibinfo {year}
  {2012})}\BibitemShut {NoStop}%
\bibitem [{\citenamefont {Mitroy}\ and\ \citenamefont
  {Zhang}(2008{\natexlab{b}})}]{pol-os}%
  \BibitemOpen
  \bibfield  {author} {\bibinfo {author} {\bibfnamefont {J.}~\bibnamefont
  {Mitroy}}\ and\ \bibinfo {author} {\bibfnamefont {J.-Y.}\ \bibnamefont
  {Zhang}},\ }\href@noop {} {\bibfield  {journal} {\bibinfo  {journal} {The
  Journal of Chemical Physics}\ }\textbf {\bibinfo {volume} {128}},\ \bibinfo
  {pages} {134305} (\bibinfo {year} {2008}{\natexlab{b}})}\BibitemShut
  {NoStop}%
\bibitem [{\citenamefont {Blundell}\ \emph {et~al.}(1991)\citenamefont
  {Blundell}, \citenamefont {Johnson},\ and\ \citenamefont
  {Sapirstein}}]{Blundell}%
  \BibitemOpen
  \bibfield  {author} {\bibinfo {author} {\bibfnamefont {S.~A.}\ \bibnamefont
  {Blundell}}, \bibinfo {author} {\bibfnamefont {W.~R.}\ \bibnamefont
  {Johnson}}, \ and\ \bibinfo {author} {\bibfnamefont {J.}~\bibnamefont
  {Sapirstein}},\ }\href {\doibase 10.1103/PhysRevA.43.3407} {\bibfield
  {journal} {\bibinfo  {journal} {Phys. Rev. A}\ }\textbf {\bibinfo {volume}
  {43}},\ \bibinfo {pages} {3407} (\bibinfo {year} {1991})}\BibitemShut
  {NoStop}%
\bibitem [{\citenamefont {Singh}\ and\ \citenamefont {Sahoo}(2014)}]{singhy}%
  \BibitemOpen
  \bibfield  {author} {\bibinfo {author} {\bibfnamefont {Y.}~\bibnamefont
  {Singh}}\ and\ \bibinfo {author} {\bibfnamefont {B.~K.}\ \bibnamefont
  {Sahoo}},\ }\href@noop {} {\bibfield  {journal} {\bibinfo  {journal} {Phys.
  Rev. A}\ }\textbf {\bibinfo {volume} {90}},\ \bibinfo {pages} {022511}
  (\bibinfo {year} {2014})}\BibitemShut {NoStop}%
\bibitem [{\citenamefont {J.~Kaur}\ and\ \citenamefont {Sahoo}(2015)}]{kaurj}%
  \BibitemOpen
  \bibfield  {author} {\bibinfo {author} {\bibfnamefont {B.~A.}\ \bibnamefont
  {J.~Kaur}, \bibfnamefont {D.~K.~Nandy}}\ and\ \bibinfo {author}
  {\bibfnamefont {B.~K.}\ \bibnamefont {Sahoo}},\ }\href@noop {} {\bibfield
  {journal} {\bibinfo  {journal} {Phys. Rev. A}\ }\textbf {\bibinfo {volume}
  {91}},\ \bibinfo {pages} {012705} (\bibinfo {year} {2015})}\BibitemShut
  {NoStop}%
\bibitem [{\citenamefont {{Safronova}}\ and\ \citenamefont
  {{Johnson}}(2008)}]{theory}%
  \BibitemOpen
  \bibfield  {author} {\bibinfo {author} {\bibfnamefont {M.~S.}\ \bibnamefont
  {{Safronova}}}\ and\ \bibinfo {author} {\bibfnamefont {W.~R.}\ \bibnamefont
  {{Johnson}}},\ }\href {\doibase 10.1016/S1049-250X(07)55004-4} {\bibfield
  {journal} {\bibinfo  {journal} {Advances in Atomic Molecular and Optical
  Physics}\ }\textbf {\bibinfo {volume} {55}},\ \bibinfo {pages} {191}
  (\bibinfo {year} {2008})}\BibitemShut {NoStop}%
\bibitem [{\citenamefont {J{\"o}nsson}\ \emph {et~al.}(2013)\citenamefont
  {J{\"o}nsson}, \citenamefont {Gaigalas}, \citenamefont {Biero{\'n}},
  \citenamefont {Fischer},\ and\ \citenamefont {Grant}}]{jonsson2013}%
  \BibitemOpen
  \bibfield  {author} {\bibinfo {author} {\bibfnamefont {P.}~\bibnamefont
  {J{\"o}nsson}}, \bibinfo {author} {\bibfnamefont {G.}~\bibnamefont
  {Gaigalas}}, \bibinfo {author} {\bibfnamefont {J.}~\bibnamefont
  {Biero{\'n}}}, \bibinfo {author} {\bibfnamefont {C.~F.}\ \bibnamefont
  {Fischer}}, \ and\ \bibinfo {author} {\bibfnamefont {I.}~\bibnamefont
  {Grant}},\ }\href@noop {} {\bibfield  {journal} {\bibinfo  {journal}
  {Computer Physics Communications}\ }\textbf {\bibinfo {volume} {184}},\
  \bibinfo {pages} {2197} (\bibinfo {year} {2013})}\BibitemShut {NoStop}%
\bibitem [{\citenamefont {Mitroy}\ and\ \citenamefont
  {Bromley}(2003)}]{mitroy2003}%
  \BibitemOpen
  \bibfield  {author} {\bibinfo {author} {\bibfnamefont {J.}~\bibnamefont
  {Mitroy}}\ and\ \bibinfo {author} {\bibfnamefont {M.}~\bibnamefont
  {Bromley}},\ }\href@noop {} {\bibfield  {journal} {\bibinfo  {journal}
  {Physical Review A}\ }\textbf {\bibinfo {volume} {68}},\ \bibinfo {pages}
  {052714} (\bibinfo {year} {2003})}\BibitemShut {NoStop}%
\bibitem [{\citenamefont {Porsev}\ and\ \citenamefont
  {Derevianko}(2006{\natexlab{b}})}]{porsev2006}%
  \BibitemOpen
  \bibfield  {author} {\bibinfo {author} {\bibfnamefont {S.}~\bibnamefont
  {Porsev}}\ and\ \bibinfo {author} {\bibfnamefont {A.}~\bibnamefont
  {Derevianko}},\ }\href@noop {} {\bibfield  {journal} {\bibinfo  {journal}
  {Journal of Experimental and Theoretical Physics}\ }\textbf {\bibinfo
  {volume} {102}},\ \bibinfo {pages} {195} (\bibinfo {year}
  {2006}{\natexlab{b}})}\BibitemShut {NoStop}%
\bibitem [{\citenamefont {Patil}(2000)}]{patil2000}%
  \BibitemOpen
  \bibfield  {author} {\bibinfo {author} {\bibfnamefont {S.}~\bibnamefont
  {Patil}},\ }\href@noop {} {\bibfield  {journal} {\bibinfo  {journal} {The
  European Physical Journal D-Atomic, Molecular, Optical and Plasma Physics}\
  }\textbf {\bibinfo {volume} {10}},\ \bibinfo {pages} {341} (\bibinfo {year}
  {2000})}\BibitemShut {NoStop}%
\bibitem [{\citenamefont {Schwerdtfeger}\ and\ \citenamefont
  {Nagle}(2019)}]{schwerdtfeger2019}%
  \BibitemOpen
  \bibfield  {author} {\bibinfo {author} {\bibfnamefont {P.}~\bibnamefont
  {Schwerdtfeger}}\ and\ \bibinfo {author} {\bibfnamefont {J.~K.}\ \bibnamefont
  {Nagle}},\ }\href@noop {} {\bibfield  {journal} {\bibinfo  {journal}
  {Molecular Physics}\ }\textbf {\bibinfo {volume} {117}},\ \bibinfo {pages}
  {1200} (\bibinfo {year} {2019})}\BibitemShut {NoStop}%
\end{thebibliography}
%

\end{document}